\documentclass[10pt,twocolumn]{article}
\pdfoutput=1
\usepackage{setspace}
 
\usepackage{graphicx}
\usepackage[utf8]{inputenc}
\usepackage{mathtools}

\usepackage{epsfig} 
\usepackage{psfrag}
\usepackage{mathrsfs}

\usepackage{multirow}

\usepackage{amsmath}
\usepackage{amssymb}
\usepackage{float}
\usepackage{amsthm}
\usepackage{graphicx}
\usepackage{color}
\usepackage{verbatim}
\usepackage{tikz,times}
\usepackage{xcolor}

\providecommand{\keywords}[1]
{
  \small	
  \textbf{\textit{Keywords---}} #1
}

\date{}

\begin{document}

\title{Lateral Force Prediction using Gaussian Process Regression for Intelligent Tire Systems}

\author{Bruno Henrique Groenner Barbosa,
        Nan Xu,
        Hassan Askari,
        Amir Khajepour
\thanks{This work has been submitted to the IEEE for possible publication. Copyright may be transferred without notice, after which this version may no longer be accessible. This research was financed in part by Coordination of Improvement of Higher Education Personnel CAPES (Grant 88881.336661/2019-01), Brazilian National Council for Scientific and Technological Development CNPq (Grant 304201/2018-9), Minas Gerais Research Foundation FAPEMIG (Grant PPM 00337/17),  National Natural Science Foundation of China (Grant Nos. 51875236 and 61790561),  China Automobile Industry Innovation and Development Joint Fund (Grant Nos. U1664257 and U1864206), and financial support of the Natural Sciences and Engineering Research Council of Canada (NSERC).
Bruno H. G. Barbosa is with Department of Automatics, Federal University of Lavras, CP 3037, 37200-000, Lavras, MG, Brazil, and Department of Mechanical and Mechatronics Engineering, University of Waterloo, N2L3G1, ON, Canada (e-mail:brunohb@ufla.br). Nan Xu is with State Key Laboratory of Automotive Simulation and Control, Jilin University, Changchun, Jilin, 130025, China, and Department of Mechanical and Mechatronics Engineering, University of Waterloo, N2L3G1, ON, Canada (e-mail:xunan@jlu.edu.cn) - Corresponding Author.
Hassan Askari and Amir Khajepour are with Department of Mechanical and Mechatronics Engineering, University of Waterloo, N2L3G1, ON, Canada (e-mail:haskari@uwaterloo.ca;a.khajepour@uwaterloo.ca).}}


\maketitle

\begin{abstract}
Understanding the dynamic behavior of tires and their interactions with road plays an important role in designing  integrated vehicle control strategies.~Accordingly, having access to reliable information about the tire-road interactions through tire embedded sensors is very demanding for developing enhanced vehicle control systems. Thus, the main objectives of the present research work are  \textit{i.} to analyze data from an experimental accelerometer-based intelligent tire acquired over a wide range of maneuvers, with different vertical loads,  velocities, and high slip angles; and \textit{ii.} to develop a lateral force predictor based on a machine learning tool, more specifically the Gaussian Process Regression (GPR) technique. It is delineated that the proposed intelligent tire system can provide reliable information about the tire-road interactions even in the case of high slip angles. Besides, the lateral forces model based on GPR can predict forces with acceptable accuracy and provide level of uncertainties that can be very useful for designing vehicle control strategies.
\end{abstract}

 


\section{Introduction}

Analyzing tire dynamics is a highly challenging task because of its complex nature and continuous interactions with roads \cite{pacejka2005tire}.~The convoluted nature of tires and their interactions with roads have attracted the attention of researchers for several years with the aim of improving vehicle safety and reducing fuel consumption \cite{nepote2005intelligent,askari2019tire,guo2007unitire,pacejka1992magic,zhao2020mssp}.~Research on the tire dynamics spans over many interesting topics, such as tire forces modeling and estimation,~contact patch identification,~tire wear effects and tire-road frictional properties.

Generally, the physics behind tire contact patch define the tire forces generations and tire-road interactions including friction and wear phenomena \cite{pottinger1992three}.~During the last two decades researchers have studied tire contact patch from different perspectives including tire design, operation conditions, tire tread pattern,  and the topography of pavement. There are three main categories of research for analyzing tire contact patch including complex mathematical modeling development, hybridization of  mathematical modeling and experimental investigation based on novel sensing systems, and intelligent tire technology \cite{askari2019tire}.

Intelligent Tires (tires with sensors) have been studied in the automotive industry since they can provide relevant vehicle dynamics information in a more reliable manner than traditional indirect estimation methods based on on-board sensors \cite{lee2019novel}.~This information can be used to improve the performance of Advanced Driver Assistance Systems (ADAS), such as the Electronic Stability Program (ESP), Traction Control Systems (TCS) and the Anti-lock Braking System (ABS), thus promoting safe driving \cite{guo2019mpc,mendoza2020strain}.

One of the most demanding parameters in tires and vehicle dynamics is the lateral force, specially when the tire experiences harsh slip angles. The case of high slip angle is where the most of estimation techniques fail to predict the tire forces with an acceptable amount of accuracy.~Estimation techniques based on machine learning algorithms have recently attracted the attention of researchers for the tire forces estimation~\cite{acosta2019virtual}.

Indeed, machine learning tools have been extensively used to identify a variety of process models \cite{aguirre2017cep,barbosaiet2019}. A widely used approach is the Gaussian Process Regression (GPR). GPR is a nonparametric probabilistic model based on Bayesian and statistical learning theories \cite{rasmussen2006}.~From this premise, GPR does not fit a specific model structure to the data underlying function. Instead, it represents the relation between inputs and outputs with theoretically infinite number of parameters, letting the data define the mapping complexity through Bayesian inference \cite{rasmussen2006}.~This makes GPR able to handle complex relations between inputs and outputs, requiring little prior knowledge of the regression problem, and using a relatively simple structure based on the mean and covariance functions \cite{SCHULZ20181}.~One of the most prominent advantages of GPR is to yield uncertainty estimates along with predictions. This feature makes it suitable for many applications such as designing model predictive control (MPC) approaches, as presented in \cite{klenske2016GPRMPCperiodic, hewing2019gprMPC}, with increasing interest in the control community \cite{ito2020kernel}.

In this paper, considering data acquired for different maneuvers with high slip angles and different velocities from an experimental accelerometer based intelligent tire system, our objective is to analyze the tire measured accelerations, understanding their relation in the tire system here described, and to develop a lateral force predictor based on GPR, which is probably the first attempt to use GPR to address this problem in the literature.~It is shown that the developed intelligent tire system paves the way for better understanding the underlying relations regarding the tire dynamics. Besides, the use of GPR is promising for tire force estimation since it provides uncertainty predictions of the estimated lateral forces that may be useful for developing more reliable vehicle control strategies. 

The remainder of this paper is organized as follows. In Section 2, we provide a literature review on the state of the art in the areas of tire contact patch identification and tire forces estimation and introduce the Gaussian Process Regression method.~A brief description of the intelligent tire system and experimental setup is provided in Section 3. The analysis of the acquired data is addressed in Section 4. The important problem of which variables to choose to compose the GPR model and its prediction results are discussed in Section 5. Final remarks and future developments are provided in Section
6.

\section{Background}

\subsection{Contact Patch and Lateral Force Estimation}

Tire contact patch shape and geometry analysis based on complex mathematical modeling was started in late 80s.~One of the earliest studies in the area of tire contact patch topology and sizes was presented by M.G. Pottinger \cite{pottinger1992three}.~In his paper, he developed a three-dimensional  mathematical model for the stress field of tire contact patch.~Wang et al. \cite{wang1994two} proposed a mathematical model to identify two-dimensional contact area of a pneumatic tire subjected to lateral forces.~The proposed mathematical model considers tire geometric parameters,~vertical deflection and the carcass camber angle.~They showed that the geometry of contact patch has a significant impact on tire transient behavior.~With employing a spring bedded ring model in the lateral, radial and circumferential directions for the tire sidewalls as well as tread rubber, Kagami et al. \cite{kagami1995analysis} developed an analytical model for tire contact patch deformation.~The proposed model was capable of predicting tire contact patch pressure distribution and camber thrust with an acceptable amount of accuracy. Nonsymmetric and more general vertical force distributions in the contact patch were studied by \cite{guo1996}, which expressed pressure distributions under different rolling resistance, inflation pressure and vertical loads conditions in a unified form.

The  combination of experimental investigation with either estimation or learning techniques for tire contact patch analysis was started  by El-Gindy et al. \cite{el1999development} in late 90s.~In their work, they analyzed tire  patch analysis based on machine learning technique.~Their trained neural network was capable of generating complex stress distribution patterns under different inflation pressure for a specific tire type.~In another research based on training a neural network,~Urbina et al. \cite{urbina2015novel} employed signals generated by twelve different sensors in a vehicle to determine tire contact patch length.~In a comprehensive research, Cossalter and Doria \cite{cossalter2005relation} studied the influence of camber angle, vertical load and inflation pressure on tire contact patch geometry.~It was shown that the size of contact patch changes with respect to the variation of camber angles and vertical loads in almost a certain fashion until the vertical load is not very high.~V. Ivanov \cite{ivanov2010analysis} used visual processing in order to estimate the tire contact patch depending on vertical loads and inflation pressure.~Researchers have also investigated the tire contact patch in truck tires. For example,~Anghelache and Moisescu \cite{anghelache2012measurement} showed the effect of inflation pressure and rolling speed on the contact patch shape and size of truck tires using a complex measurement system, which includes 30 sensing elements located in the road.~It was represented that the contact patch size decreases with increasing the inflation pressure and rolling speed in the case of free rolling.~In an interesting research, Xiong and Tuononen \cite{xiong2014laser} developed an optical sensing system for tire contact patch analysis with focus on tire tread deformation. Their experimental results clearly reveal the asymmetric deformation of tire tread owing to rolling resistance. 

The use of intelligent tire for contact patch length estimation was experimentally studied by Khaleghian et al. \cite{khaleghian2016estimation} in 2015.~Based on the data generated by an accelerometer, which was attached to the inner liner of a tire, they estimated tire contact patch length using the two consecutive peaks in accelerometers signals.~Lee and Taheri \cite{lee2019novel} used the technology of intelligent tire in combination with flexible ring tire model to identify tire contact patch shape, stiffness of the wall and sidewall, and contact patch distribution.~Very recently, based on the technology of intelligent tire, Petit et al. \cite{mendoza2020strain} identified tire effective rolling radius and contact length.~Employing the measurements obtained from a strain-based sensor inside the tire and fuzzy logic system enabled them to accurately estimate tire contact patch length for a few simple driving conditions.

As stated in the introduction section, tire forces estimation is another important path of research in the area of tire dynamics. In fact, the estimation of these forces is a challenging task in the area of vehicle dynamics, especially when a vehicle experiences harsh maneuver \cite{baffet2009estimation,viehweger2020vehicle,singh2018literature}.~Having an accurate estimation from tire forces can be hugely effective in vehicle stability and handling performance. During the last two decades, several researchers have either developed or utilized both traditional and advanced estimation techniques to obtain tire forces in an online fashion \cite{guo2019review,cho2009estimation,hashemi2017corner,rezaeian2014novel,ray1997nonlinear}.~With the implementation of rigorous estimation techniques, they have been able to obtain tire forces in some extent, however it is still needed to develop more reliable approaches to capture tire forces with high accuracy in all possible driving maneuvers. Generally, two categories of estimations are developed in vehicle systems.  These are dependent on the modeling schemes and are categorized as either kinematic or dynamic model.~The estimation techniques based on these two categories are fully connected to tire models, which results in estimation bias due to incorrect tire and road parameters \cite{doumiati2012vehicle}. Owing to the above-mentioned shortcomings of estimation techniques,~very recently,~the implementation of machine learning techniques in conjunction with direct measurement from embedded sensors in tires have attracted the attention of researchers for the online prediction of tire forces \cite{khaleghian2019combination,askari2019towards}, which is the main theme of the present article.

\subsection{Gaussian Process Regression}

In order to understand how GPR works, let $f$ represent an unknown function mapping the input vector $\mathbf{x}$ to output $y$ ($f: \mathbf{x} \rightarrow y$); and $n$ training independent and identically distributed data points are available from some unknown distributions, $\{ \mathbf{x}_i, y_i \}$, with $ i= 1\ldots n$, where $\mathbf{x}_i \in \mathbb{R}^r$ and $y_i \in \mathbb{R}$. Assume that the output is a function of the input vector, such that 

\begin{equation}
   y_i = f(\mathbf{x}_i) + \epsilon_i, 
\end{equation}

\noindent where $\epsilon_i$ is the i.i.d noise term, which follows a normal distribution, $\epsilon \sim \mathcal{N}(0, \sigma_\epsilon^2$). 

In a GPR, the function $f(\mathbf{x})$ is distributed as a Gaussian process, a set of random variables where any finite number of which have a joint (multivariate) Gaussian distribution. That is, as a distribution over functions:

\begin{equation}
    f(\mathbf{x}) \sim \mathcal{GP}(\mu(\mathbf{x}),\kappa(\mathbf{x},\mathbf{x}')),
\end{equation}

\noindent which is defined by a mean and a covariance functions, given, respectively, as
\begin{align}
    \mu(\mathbf{x}) &= \mathrm{E}[f(\mathbf{x})], \\
    \kappa(\mathbf{x},\mathbf{x}') & = \mathrm{E}[(f(\mathbf{x}-\mu(\mathbf{x}))(f(\mathbf{x}'-\mu(\mathbf{x}'))],
\end{align}
 
 \noindent where the prior of $\mu(\mathbf{x})$ is taken as zero herein, as usual, to avoid computations and to perform inference by means of the covariance function only \cite{SCHULZ20181}. The covariance function, normally called $kernel$, controls the smoothness and defines similarity between data points, which is crucial for GPR \cite{rasmussen2006}. The choice of an appropriate kernel is very important and many covariance functions have been proposed, such as the squared exponential and radial basis function, which is a special case of the Matérn kernel. The Matérn covariance between two points with distance $\tau = |\mathbf{x} - \mathbf{x}'|$ is expressed as \cite{rasmussen2006}:
 
 \begin{equation}
 \label{eq:matern}
     \kappa_v(\tau) = \frac{2^{1-v}}{\Gamma(v)}\left( \sqrt{2v}\frac{\tau}{l}\right )^v K_v \left(\sqrt{2v} \frac{\tau}{l}\right ),
 \end{equation}
 
 \noindent where $\Gamma$ is the gamma function,~$K_v$ the modified Bessel function, with $v$ and $l$ as positive parameters.
 
 After defining the covariance function, the prior distribution of the target values can be described by
 
 \begin{equation}
     \mathbf{y} \sim \mathcal{N} (\mathbf{0},\mathbf{K}(\mathbf{X},\mathbf{X}) + \sigma_\epsilon \mathbf{I}),
 \end{equation}
 
 \noindent where $\mathbf{K}(\mathbf{X},\mathbf{X}) \in \mathbb{R}^{n \times n}$ is the covariance matrix between all observed training points and $\mathbf{I}$ is the identity matrix.

In order to make prediction for a new test input vector $\mathbf{x_*}$ by drawing $f_*$, and resembling that the training outputs $\mathbf{y}$ and prediction $f_*$ follow a joint (multivariate) Gaussian distribution, as below: 

\begin{align}
\begin{bmatrix}
\mathbf{y}\\
f_*
\end{bmatrix} \sim & \mathcal{N} \begin{pmatrix}  \mathbf{0}, \begin{bmatrix}
\mathbf{K}(\mathbf{X},\mathbf{X}) + \sigma_\epsilon \mathbf{I} & \mathbf{K}(\mathbf{X},\mathbf{x_*})\\
\mathbf{K}(\mathbf{X},\mathbf{x_*})' & \mathbf{K}(\mathbf{x_*},\mathbf{x_*})
\end{bmatrix} \end{pmatrix}, 
\end{align}
\vspace{0.1cm}

\noindent where $\mathbf{K}(\mathbf{x_*},\mathbf{x_*})$ is the covariance matrix between the new input vector values, and $\mathbf{K}(\mathbf{X},\mathbf{x_*})$ is the covariance between training and test inputs, thus, the posterior distribution of the prediction $f_*$ at a test point $\mathbf{x_*}$ is a multivariate normal distribution given by

\begin{equation}
    P(f_*|\mathbf{X},\mathbf{y},\mathbf{x_*}) \sim \mathcal{N}({\mu}_*,{\sigma}_{f_*}^2),
\end{equation}

\noindent where the mean and variance of $f_*$ are, respectively, expressed as 
\begin{align}
    {\mu}_* & = \mathbf{K}(\mathbf{x_*},\mathbf{X})[\mathbf{K}(\mathbf{X},\mathbf{X}) + \sigma_\epsilon \mathbf{I} ]^{-1} \mathbf{y}, \label{eq:gpr1}\\
    {\sigma}_{f_*}^2 & = \mathbf{K}(\mathbf{x_*},\mathbf{x_*}) - \mathbf{K}(\mathbf{x_*},\mathbf{X}) [\mathbf{K}(\mathbf{X},\mathbf{X}) + \sigma_\epsilon \mathbf{I} ]^{-1} \mathbf{K}(\mathbf{x_*},\mathbf{X}).\label{eq:gpr2}
\end{align}

Thus, new outputs can be estimated as the mean values of the aforementioned conditional distribution. It is worth mentioning that the variance of the provided output is expected to reduce when the number of training samples increases, reflecting the decreasing of the estimation uncertainty.~Providing the prediction variance is an interesting feature of GPR that distinguishes it from other machine learning methods \cite{jin2015}, and makes it a promising candidate for tire lateral force prediction.

\section{The Intelligent Tire System and the Experimental Data}
\label{sec:methods}

\begin{figure*}[!h]
 \centering
  \includegraphics[width=0.65\textwidth]{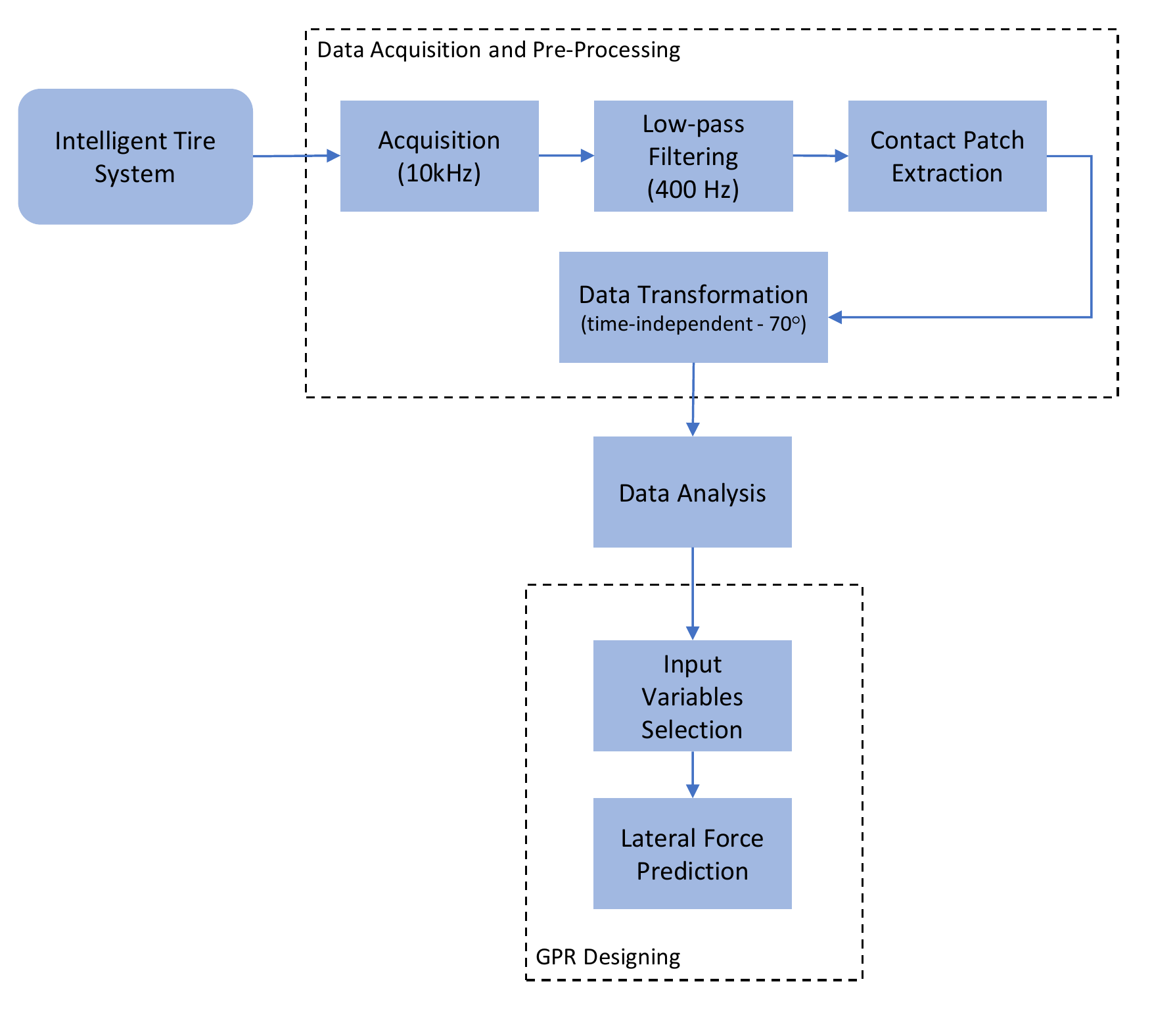}
  \caption{Flowchart of the methodology steps.}
 \label{fig:method}
\end{figure*}

In order to analyze the intelligent tire system data and to develop the lateral force prediction model based on machine learning tools, the methodology steps presented in Fig.~\ref{fig:method}, are implemented. Firstly, the test bench composed of sensors, actuators and a data acquisition device is introduced. Then, the data acquisition and pre-processing steps are explained followed by the experimental setup. Later, data analysis and force prediction steps are described in Sections \ref{sec:analysis} and \ref{sec:force}.

\subsection{The Intelligent Tire system}
\label{sec:probStat}

\begin{figure}[htb]
 \centering
  \includegraphics[width=.2\textwidth]{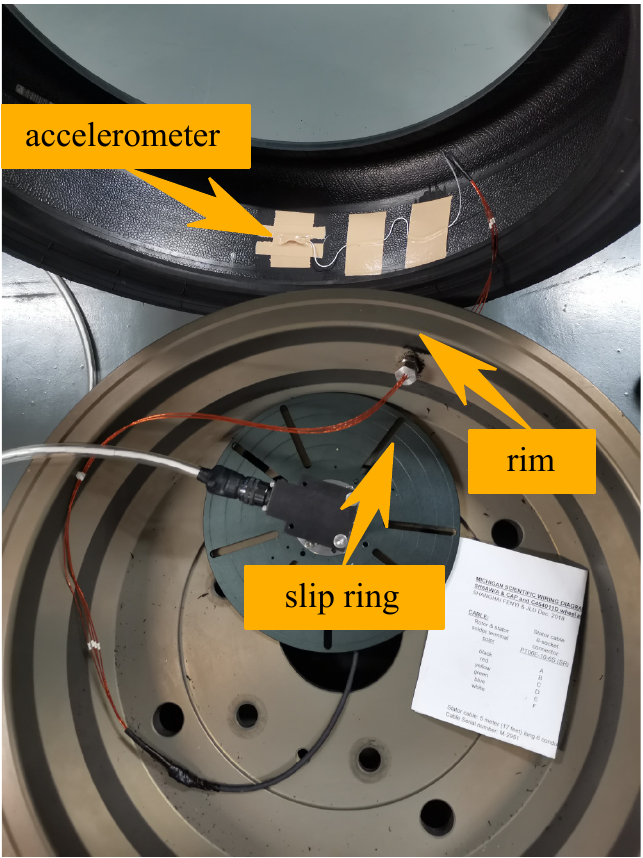}
  \caption{Intelligent tire system.}
  \label{fig1}
\end{figure}

The intelligent tire system studied in this work consists of a tri-axial acceleration sensor, a slip ring, a signal regulator, a NI data acquisition system (DAQ).~In the current setup, the tri-axial acceleration sensor is
attached and fixed at an arbitrary point along the central inner liner of the tire as shown in Fig. \ref{fig1}, and measures the longitudinal, lateral, and vertical accelerations in the $x$, $y$ and $z$ directions of its body coordinate frame, respectively. A slip-ring device is mounted to the rim, which transmits the sensor signals from the rotating tire to the DAQ system through the signal regulator. The signal debugger provides energy supply for the signal, while the NI acquisition system can adjust signal channel and sampling frequency to collect the acceleration signals.

\begin{figure*}[htb]
 \centering
  \includegraphics[width=.6\textwidth]{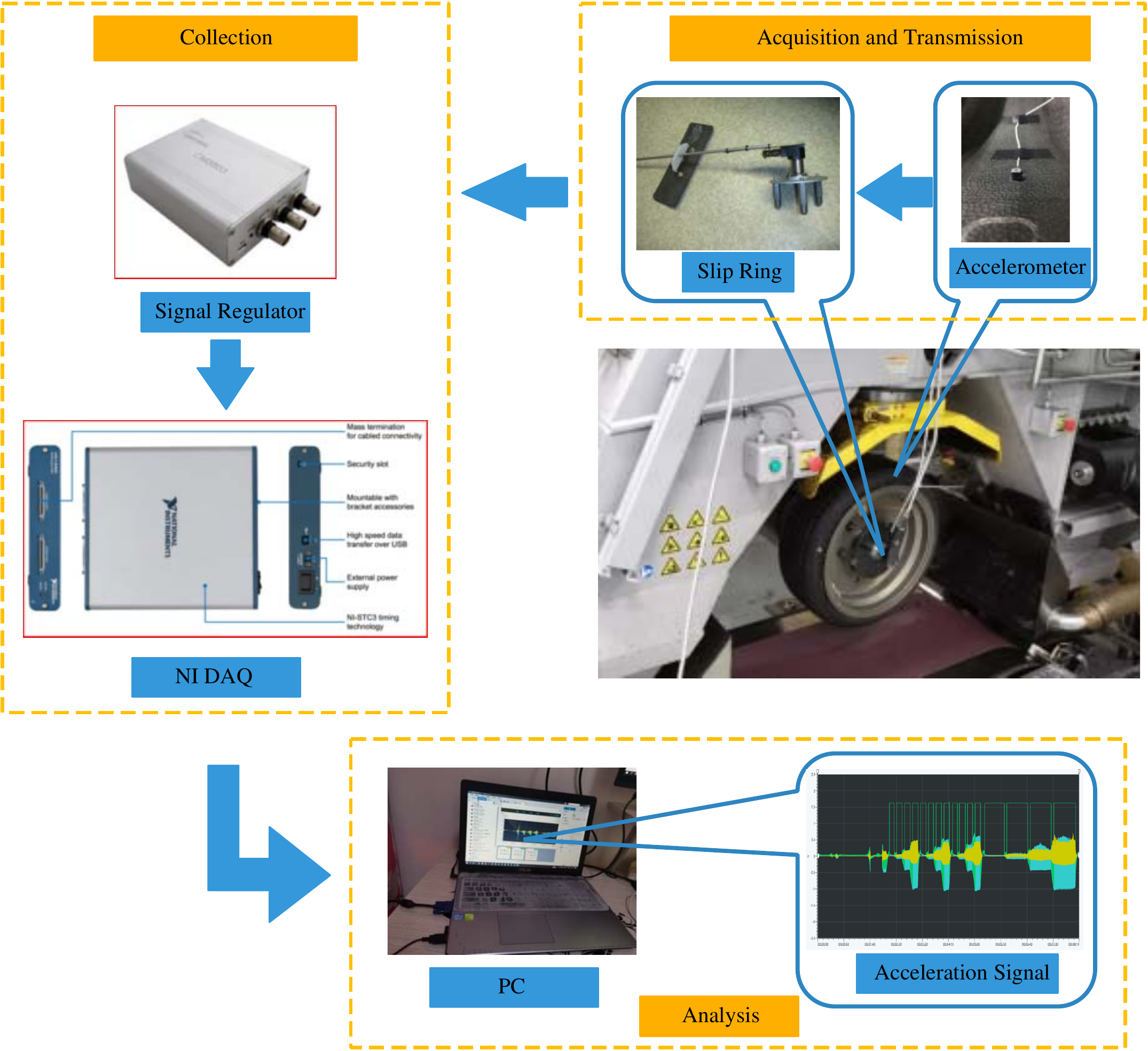}
  \caption{The entire testing system.}
  \label{fig2}
\end{figure*}

To acquire the acceleration data under conditions with different tire slip angles, the MTS Flat-Trac testing system is used. The required vertical load, driving torque, speed, and tire slip angle can be applied to simulate different  conditions for tires.~At the same time, the acceleration signals are recorded by the NI DAQ system.~In
addition, the slip angles and tire forces in six directions can be measured and recorded by the DAQ system of the MTS testbed. The diagram showing the whole testing system is presented in Fig. \ref{fig2}.

\subsection{Data Acquisition and Pre-processing}

\begin{figure}[htb]
\centering
 \includegraphics[width=.75\linewidth]{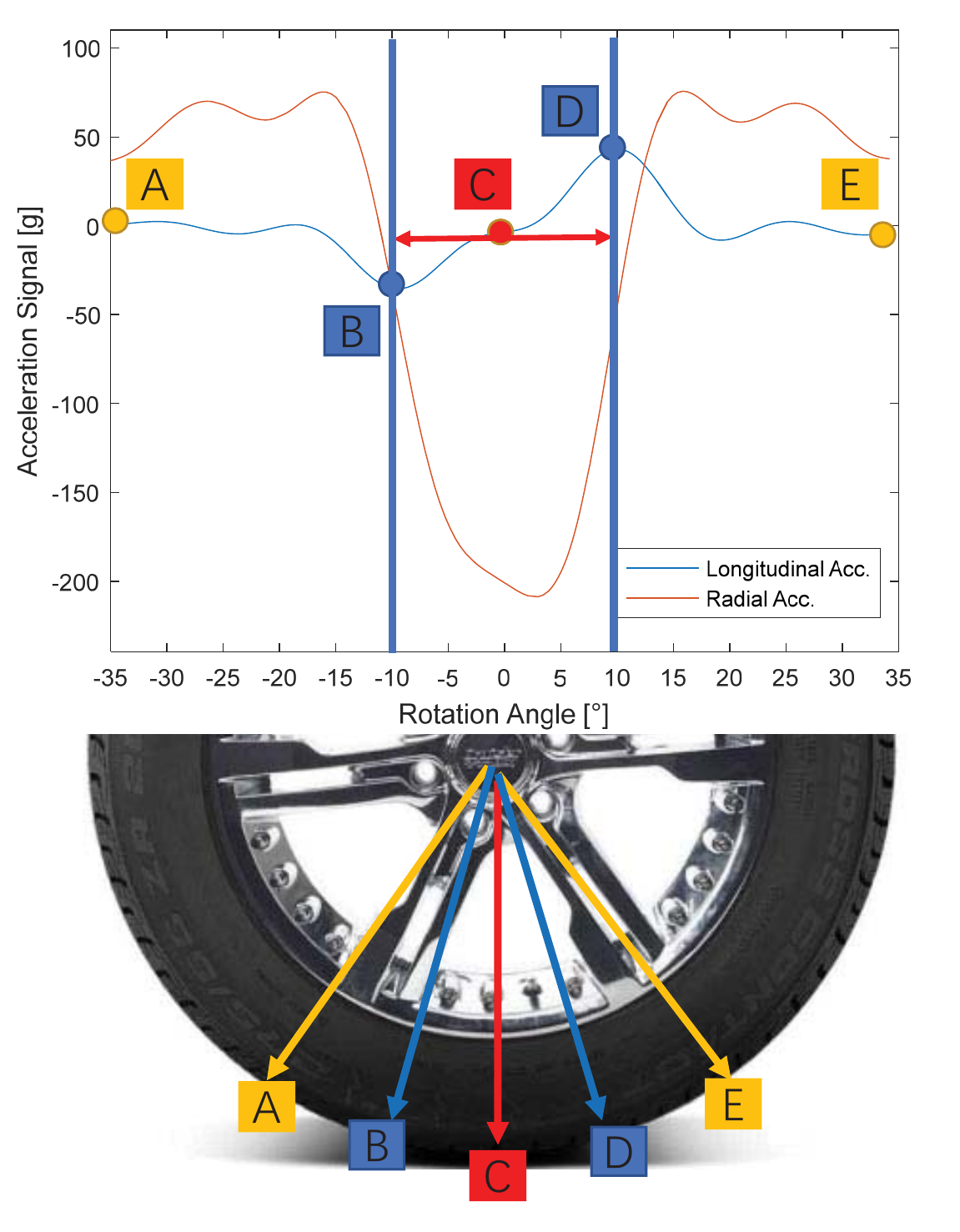}
\caption{Contact patch and the featured region extraction.}
 \label{fig:contacpatch}
\end{figure}

In order to analyze the acceleration measurements and to develop the lateral force soft sensor, only acceleration data acquired adjacent to and inside the contact patch are useful and the acceleration in other places may be neglected. In this way, the raw data collected at the sampling frequency of 10kHz must be pre-processed, and the whole process includes data filtering, selection and transformation, as shown in Fig.~\ref{fig:method}. The data pre-processing procedure is as follows:  
\begin{enumerate}
	\item \textit{Low-pass filtering}:~a low-pass Butter-worth filter with order 5 and cut-off frequency of 400Hz is used to filter  the collected acceleration signals.
	\item \textit{Contact patch extraction}: two peaks can be observed in the circumferential direction (longitudinal acceleration - $Ac_x$) when the accelerometer enters and leaves the contact patch. Therefore, after extracting data for a full tire revolution by means of the encoder signal, the contact patch data can be also obtained. As shown in Fig.~\ref{fig:contacpatch}, the points B and D can be considered as the starting and ending positions of the contact patch.
	\item \textit{Data transformation}: in order to make the contact patch data with the same number of samples, \textit{i.~e.} independent of the tire rolling speed, time-dependent measurements are processed, as presented in Fig.~\ref{fig:contacpatch}. Specifically, by using the encoder signal and the current tire rolling speed, the acceleration for any point along the inner liner can be found. After identifying the contact patch in step 2, the center of contact patch C can be determined as shown in Fig. \ref{fig:contacpatch}. For any testing condition, the angle range of the contact patch is less than $ {70}^{\circ}$. Then, from the center point C, point A and E are found with a whole angle of ${70}^{\circ}$. Finally, one single data per each ${0.5}^{\circ}$ within point A and E is extracted, yielding a total of 140 points independent of the tire speed. In this way, the acceleration data, in  three directions - longitudinal ($Ac_x$), lateral ($Ac_y$) and radial ($Ac_z$) - are always from the same tire location. 
	
\end{enumerate}

\begin{figure*}[!htb]
\centering
\begin{tabular}{cc}
(a) & (b) \\
\includegraphics[width=0.4\linewidth]{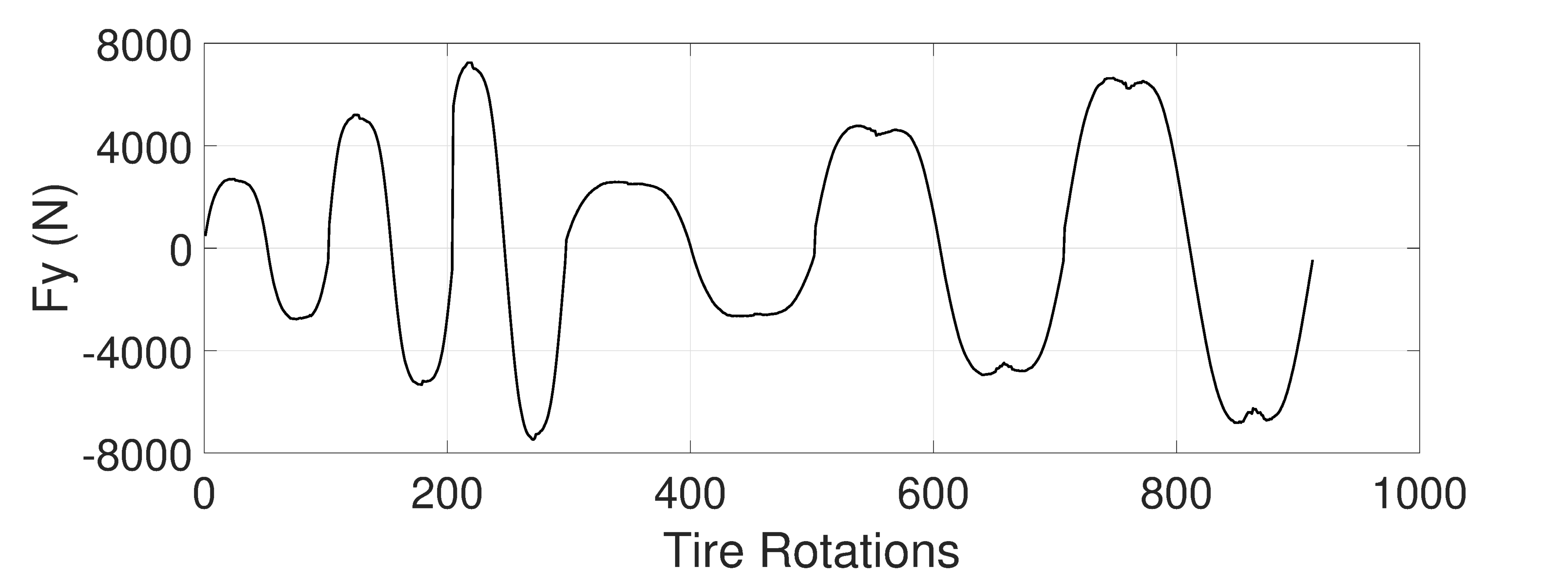} & \includegraphics[width=0.4\linewidth]{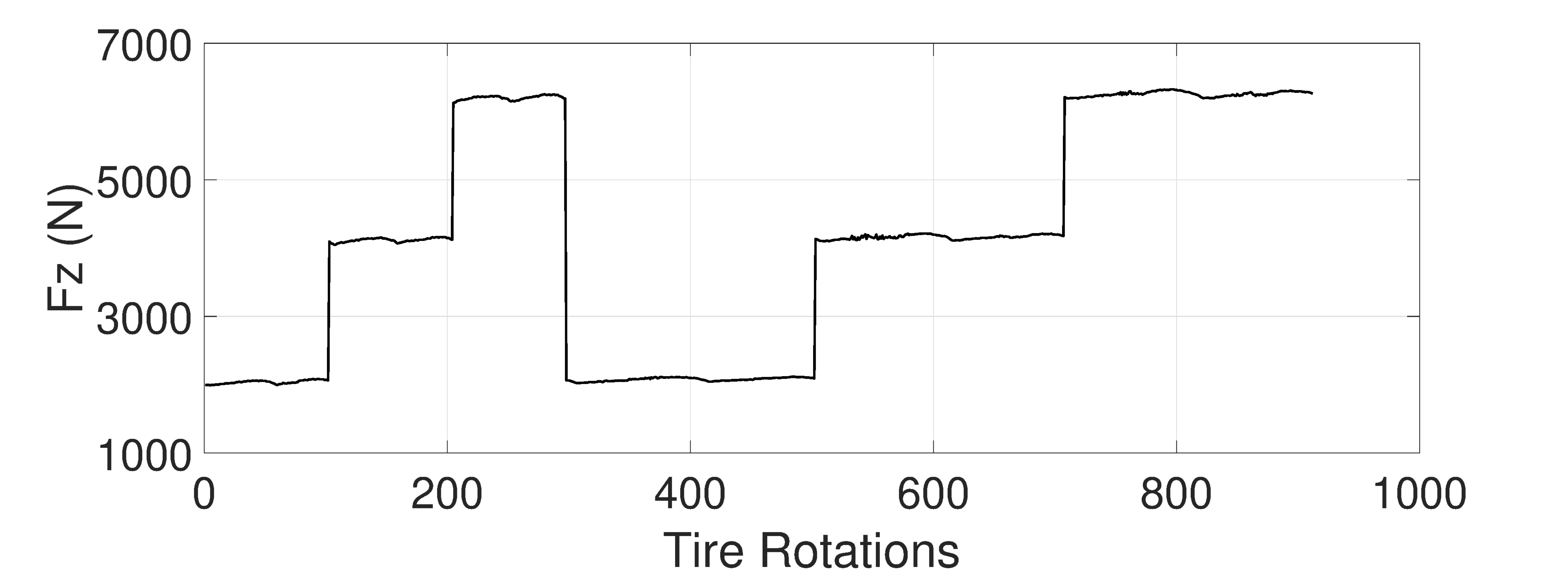} \\
(c) & (d) \\
\includegraphics[width=0.4\linewidth]{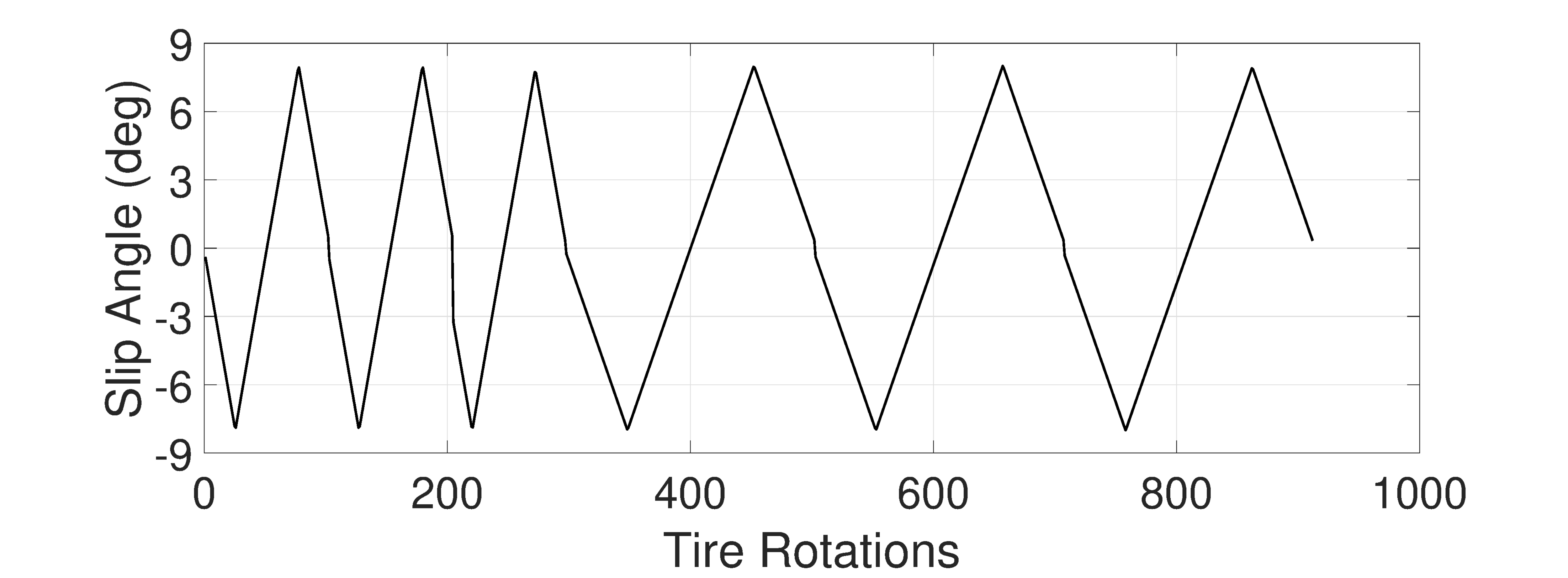} &
\includegraphics[width=0.4\linewidth]{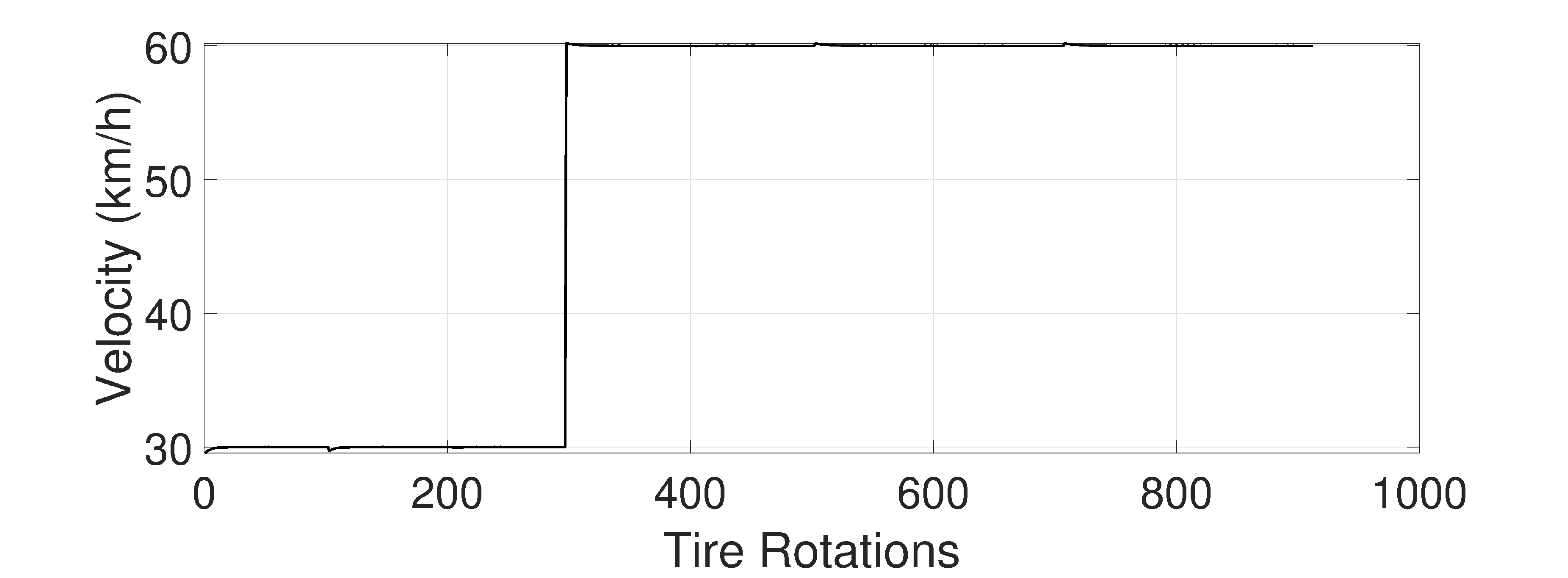}
\end{tabular}
\caption{Measurements over tire rotations of Data set 1.}
\label{fig:dataset_triang}
\end{figure*}

\begin{figure*}[!htb]
\centering
\begin{tabular}{cc}
(a) & (b) \\
\includegraphics[width=0.4\linewidth]{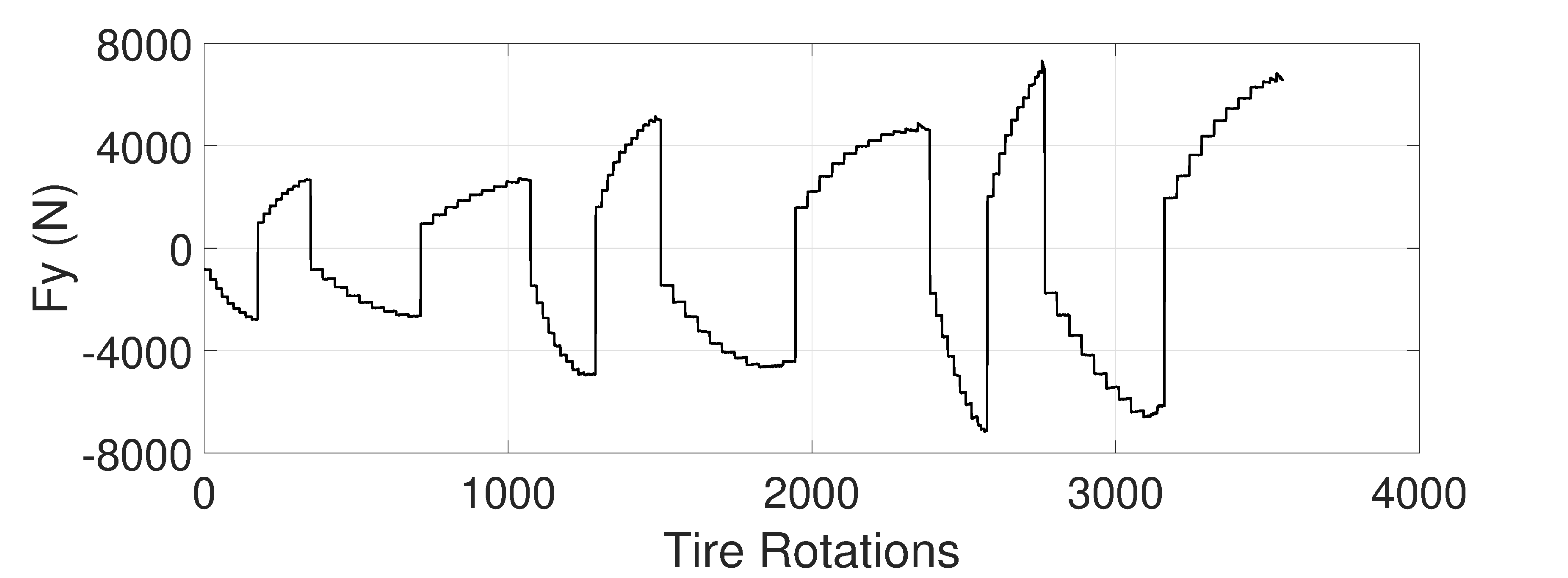} & 
\includegraphics[width=0.4\linewidth]{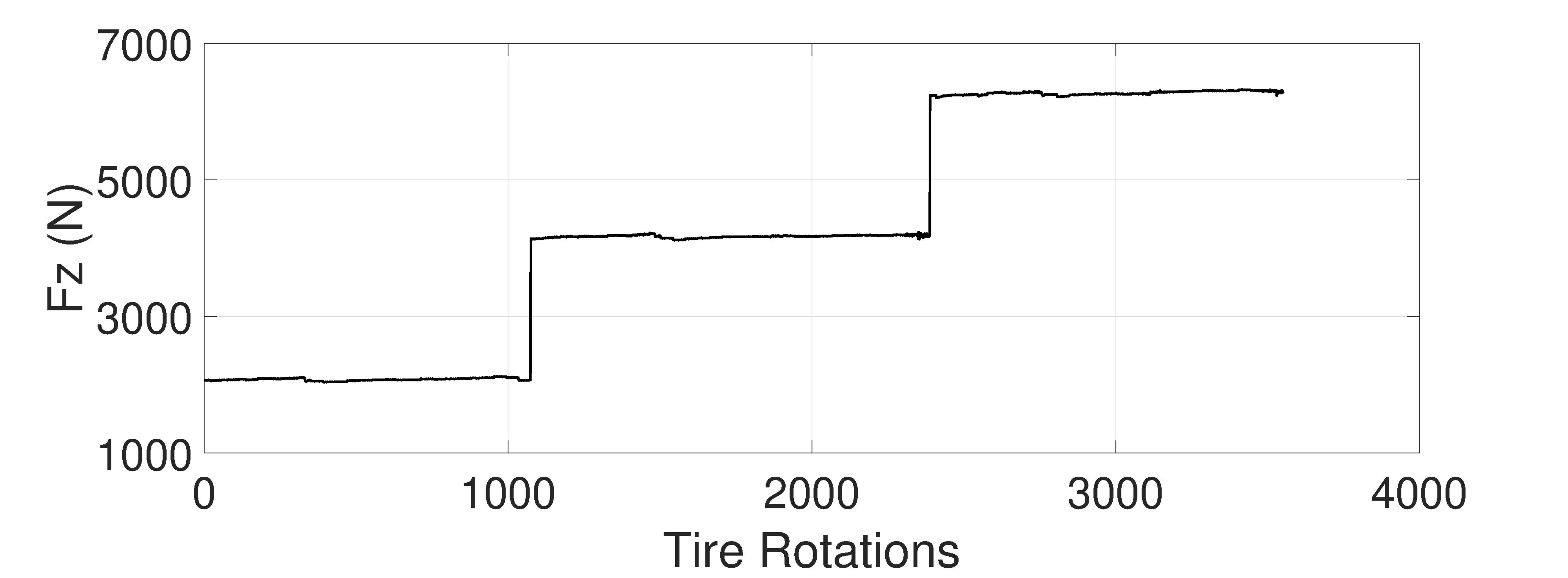} \\
(c) & (d) \\
\includegraphics[width=0.4\linewidth]{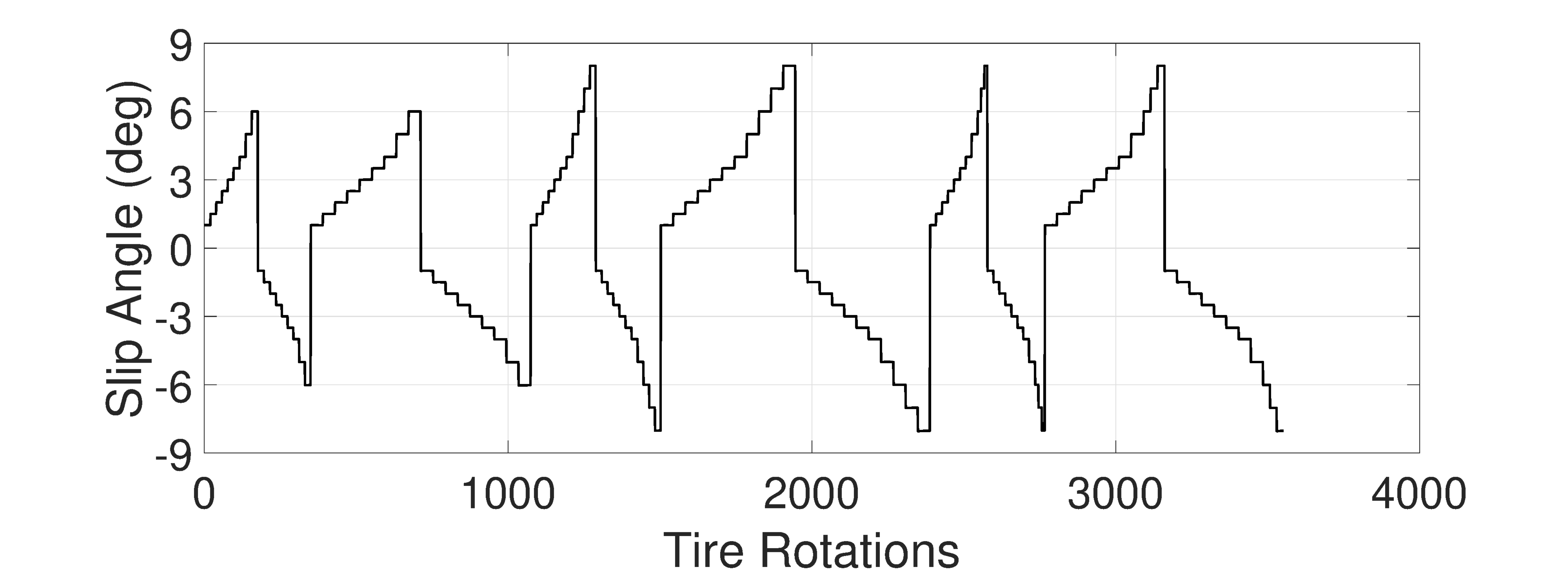} &
\includegraphics[width=0.4\linewidth]{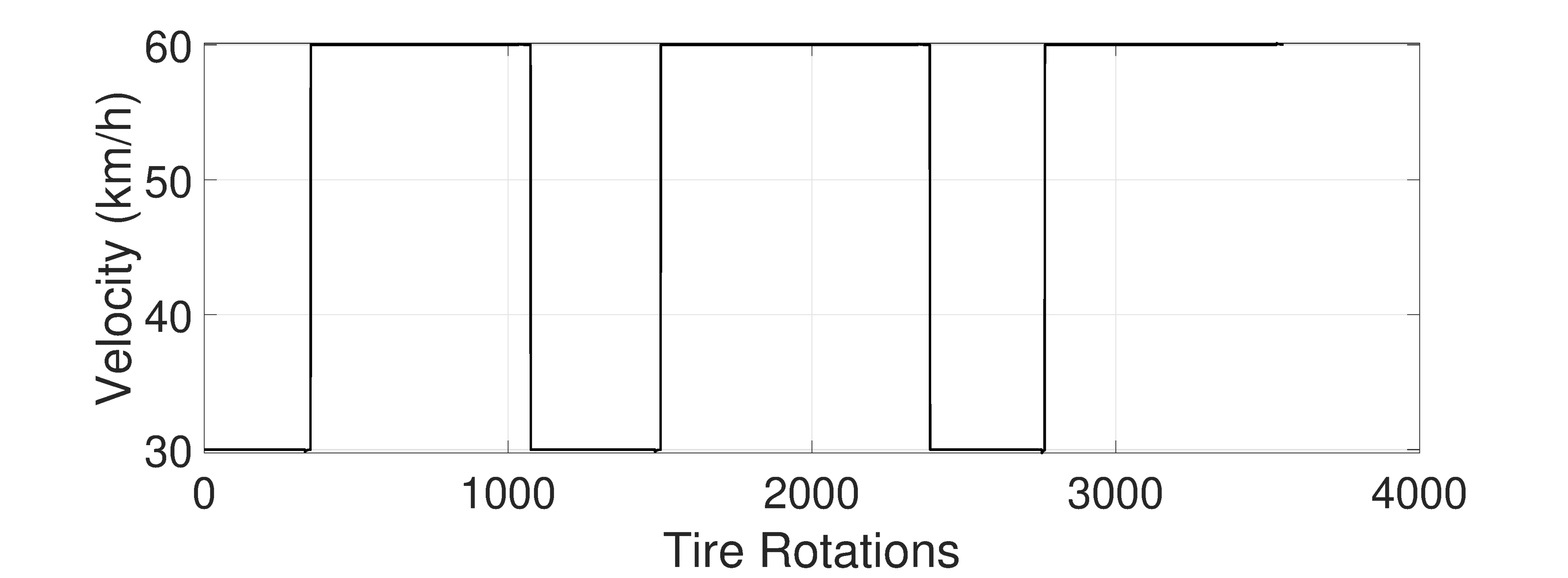}
\end{tabular}
\caption{Measurements over tire rotations of Data set 2.}
\label{fig:dataset_step}
\end{figure*}

To properly study the measured accelerations and to develop the lateral force soft sensor, a wide range of maneuvers were conducted with different vertical loads, velocities and slip angles, as shown in Table \ref{Table1}. In total, the data consists of 4464 tire rotations (912 for Data set 1 and 3552 for Data set 2), as shown in Fig.~\ref{fig:dataset_triang} and Fig.~\ref{fig:dataset_step}. The main difference between the data sets is how the slip angle varies during the experiments. As can be inferred from Fig.~\ref{fig:dataset_triang} (c), the slip angle varies continuously over the tire rotations in a triangular wave manner for Data set 1. Nonetheless, in Data set 2, the slip angle is kept constant during some tire rotations for specific values as shown in Fig.~\ref{fig:dataset_step} (c).     

\begin{table}[htb]
	\centering
	\caption{Testing scenarios}
	\label{Table1}
	\begin{tabular}{cc}
		
		\textbf{Parameters} & \textbf{Values}  \\ \hline
		Pressure (kPa) & 220   \\ \hline
		Vertical Load (N) & 2080, 4160, 6240  \\ \hline
		Velocity (km/h) & 30, 60 \\ \hline
		Slip Angle range (deg) & $\pm8$ \\ \hline
	
	\end{tabular}
\end{table}

\section{Tire Acceleration Measurements Analysis}
\label{sec:analysis}

\begin{figure*}[!ht]
\centering
\begin{tabular}{cc}
(a) & (b) \\
\includegraphics[width=0.4\linewidth]{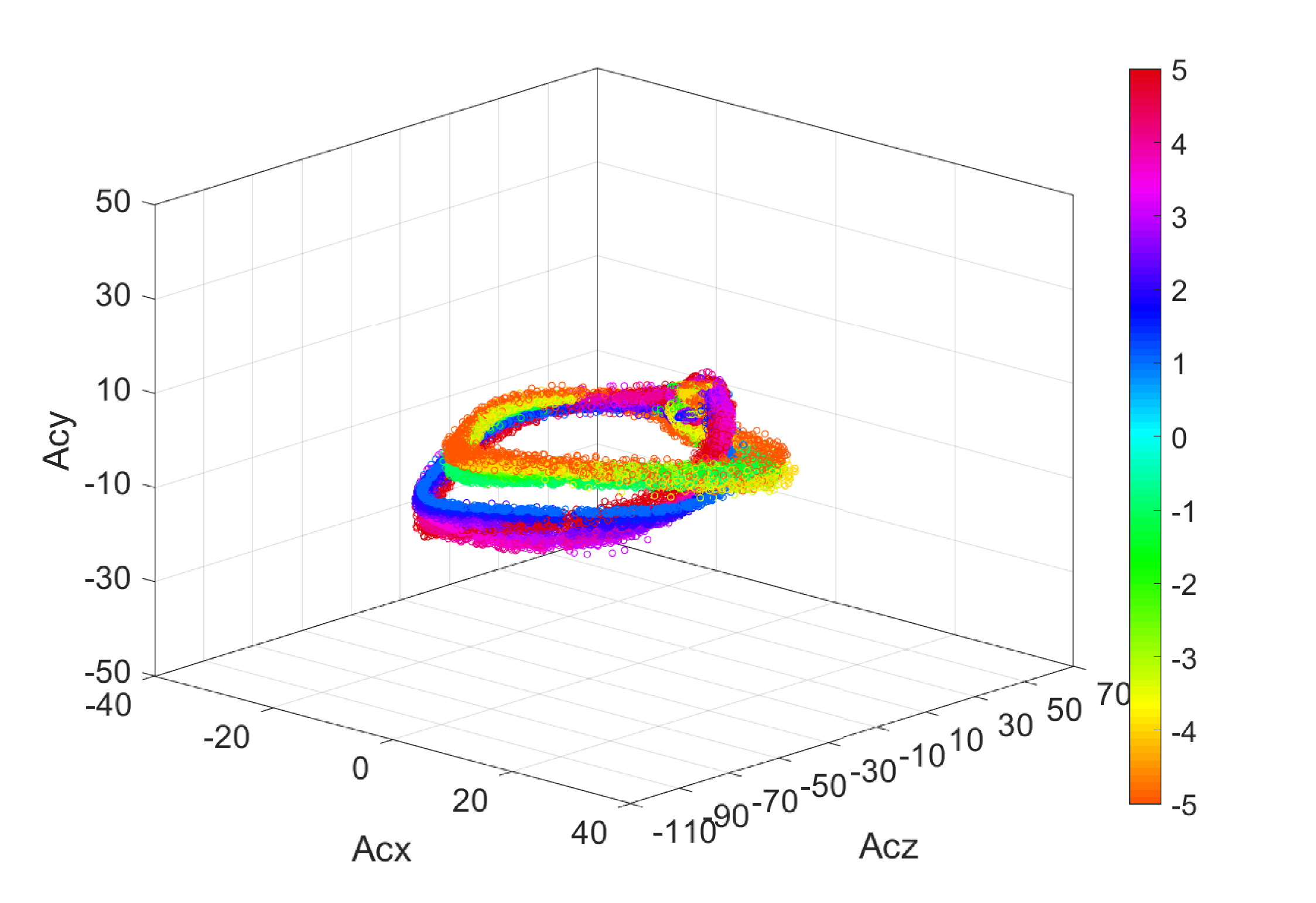} & 
\includegraphics[width=0.4\linewidth]{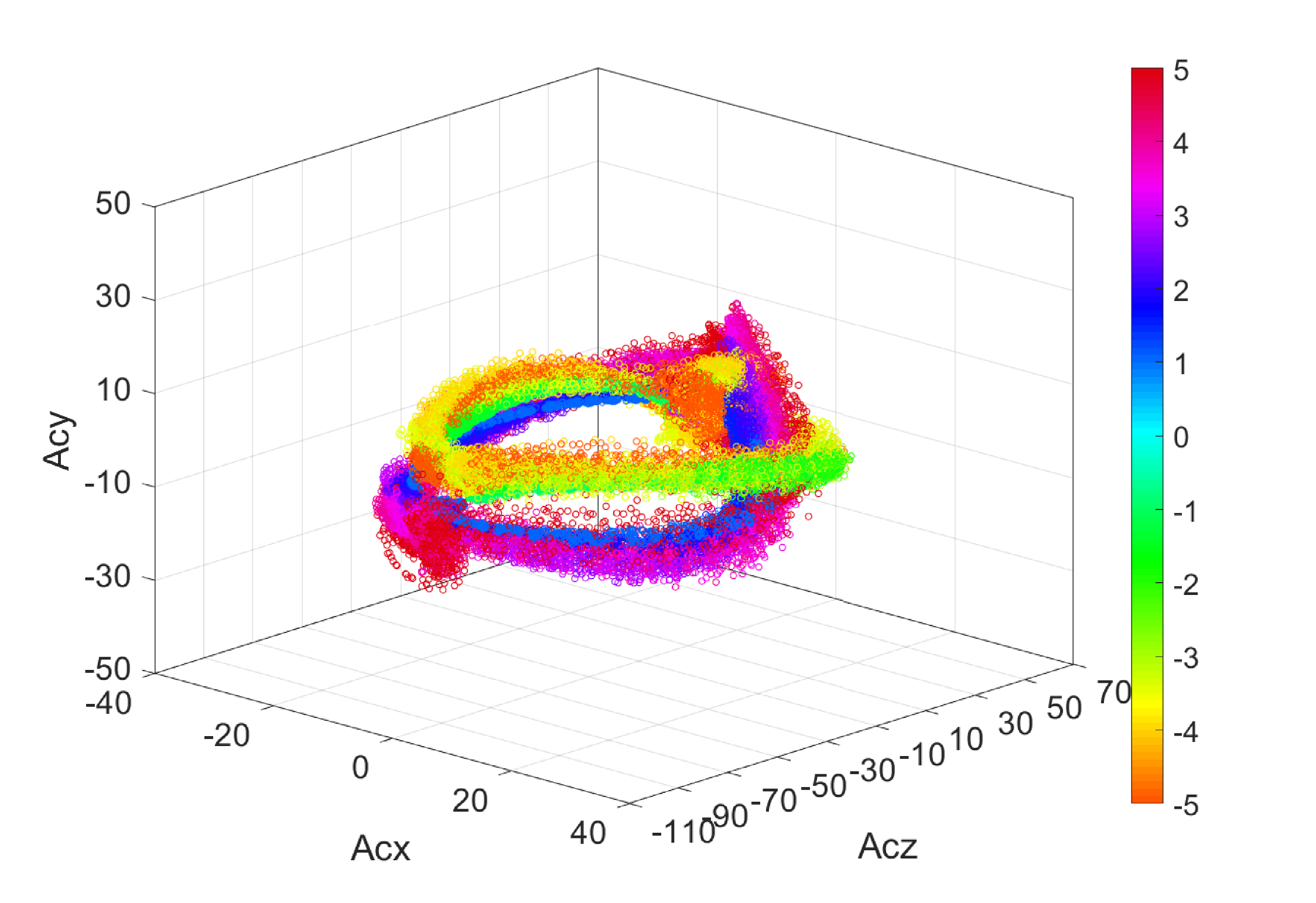}
\end{tabular}
\begin{tabular}{c}
(c) \\
\includegraphics[width=0.4\linewidth]{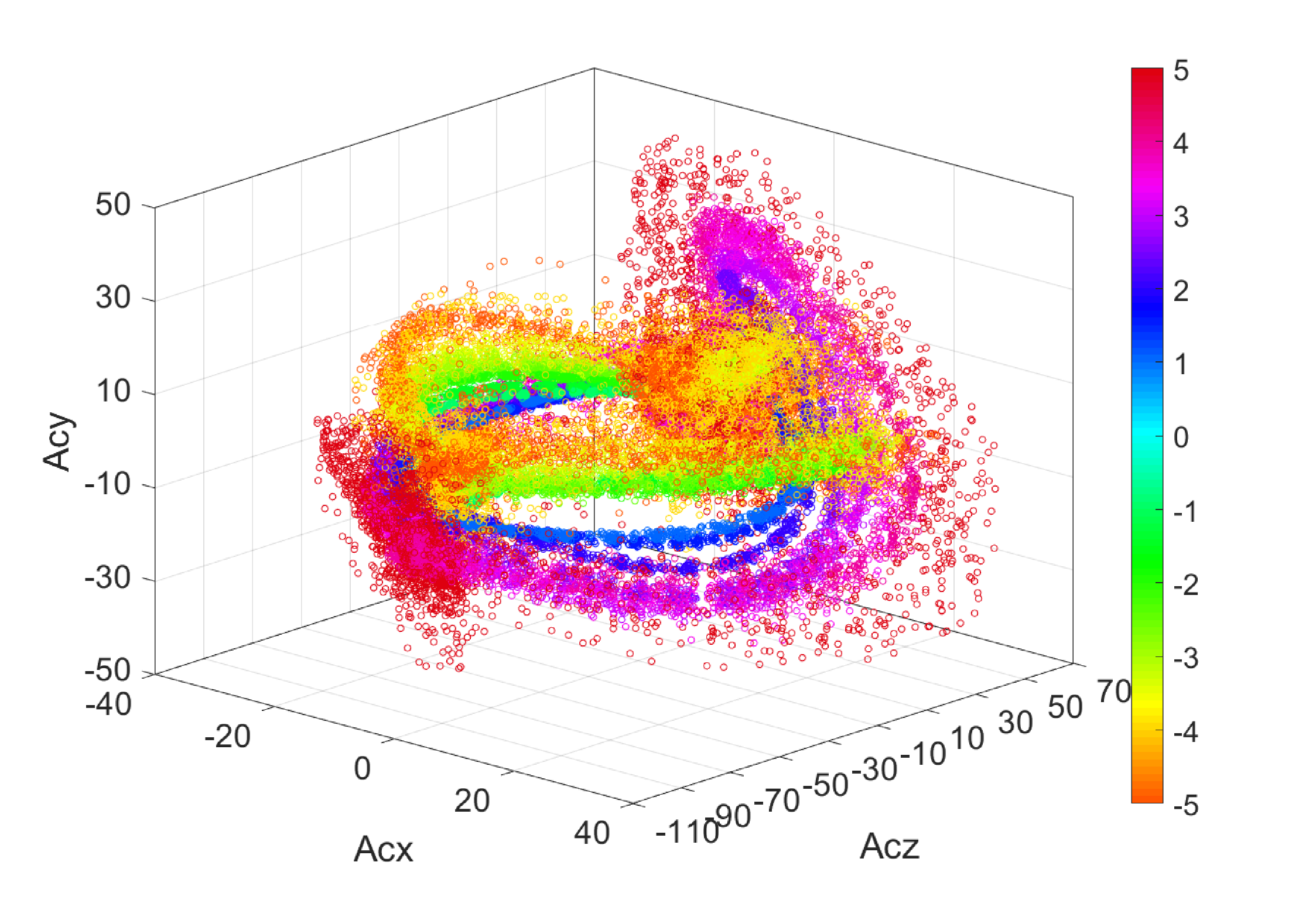} 
\end{tabular}
\caption{Samples of accelerations measured over high tire speed experiments, for different loads and slip angles (Data set 2): (a) $F_z = 2080 $ N, (b) $F_z = 4160$~N and (c)  $F_z = 6240$ N. Colors represent different slip angles.}
\label{fig:axyz_3D}
\end{figure*}

In this section, the  acceleration signals in different directions $(x,y,z)$ for several testing conditions are presented. Figure \ref{fig:axyz_3D} (a-c) represent the correlation of acceleration signals in high speed testing cases and three different normal loads. The color-bar represents the range of slip angles.~A very clear border can be found between the negative and positive slip angles, especially in the case of lower normal load. As the normal load increases the scattering region of acceleration components expands.~Thus, it shows that acceleration signals are directly connected to the normal loads as all of them are increased by increasing the normal loads. This behavior in acceleration signals resembles the correlation of longitudinal and lateral forces to normal loads.

\begin{figure*}[!htb]
\centering
\begin{tabular}{cc}
\includegraphics[width=0.4\linewidth]{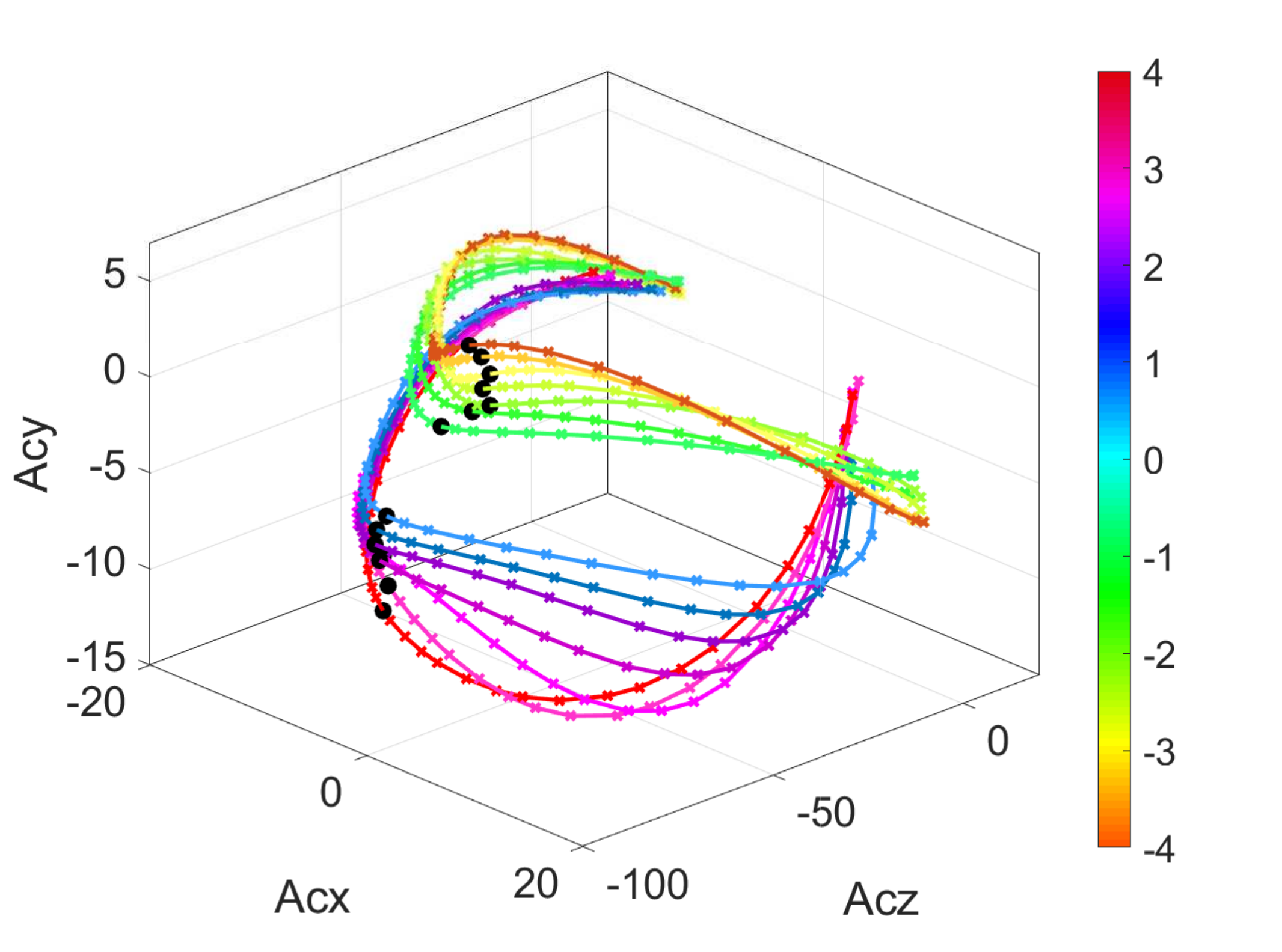} & 
\includegraphics[width=0.4\linewidth]{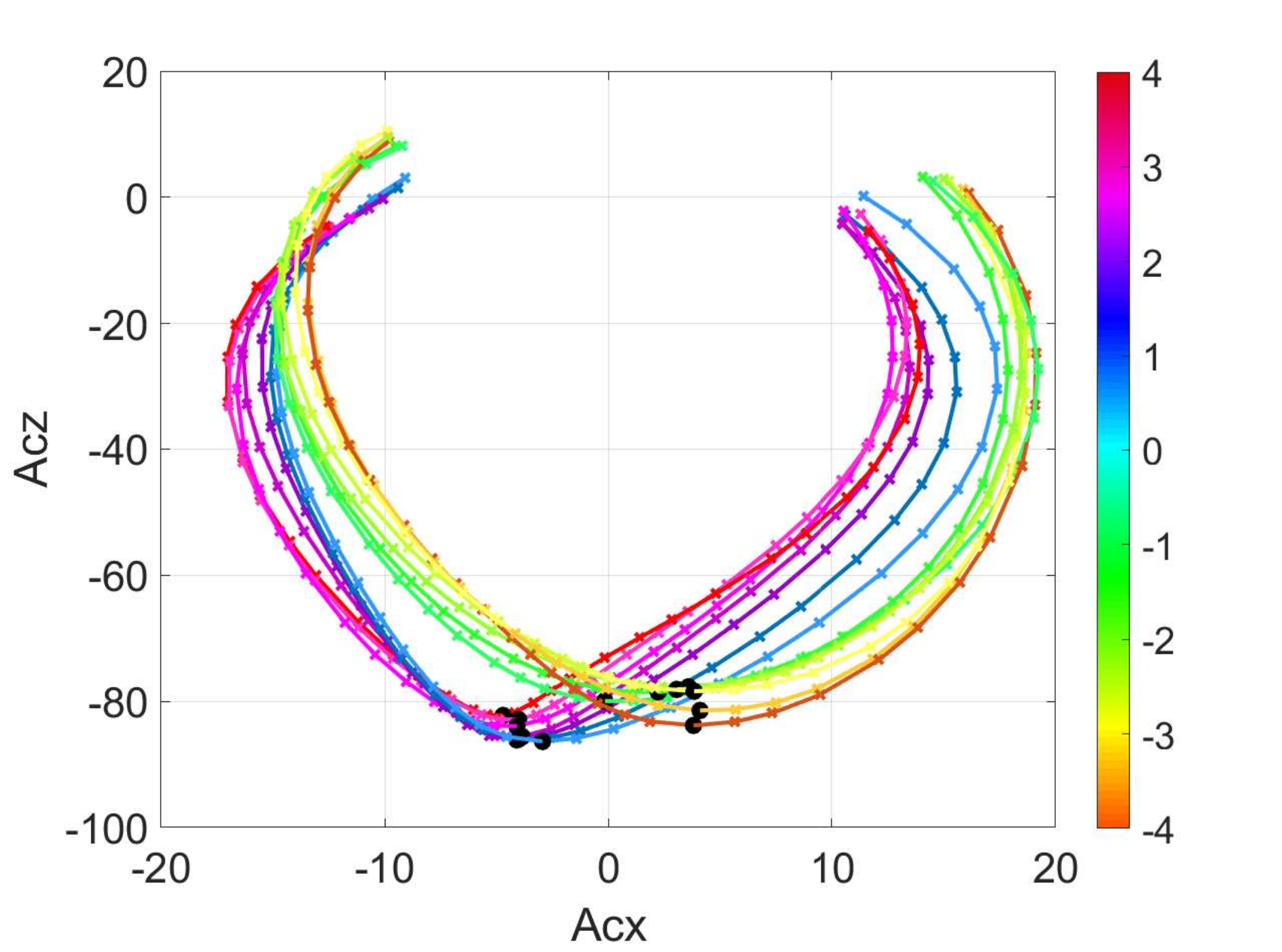} \\
\includegraphics[width=0.4\linewidth]{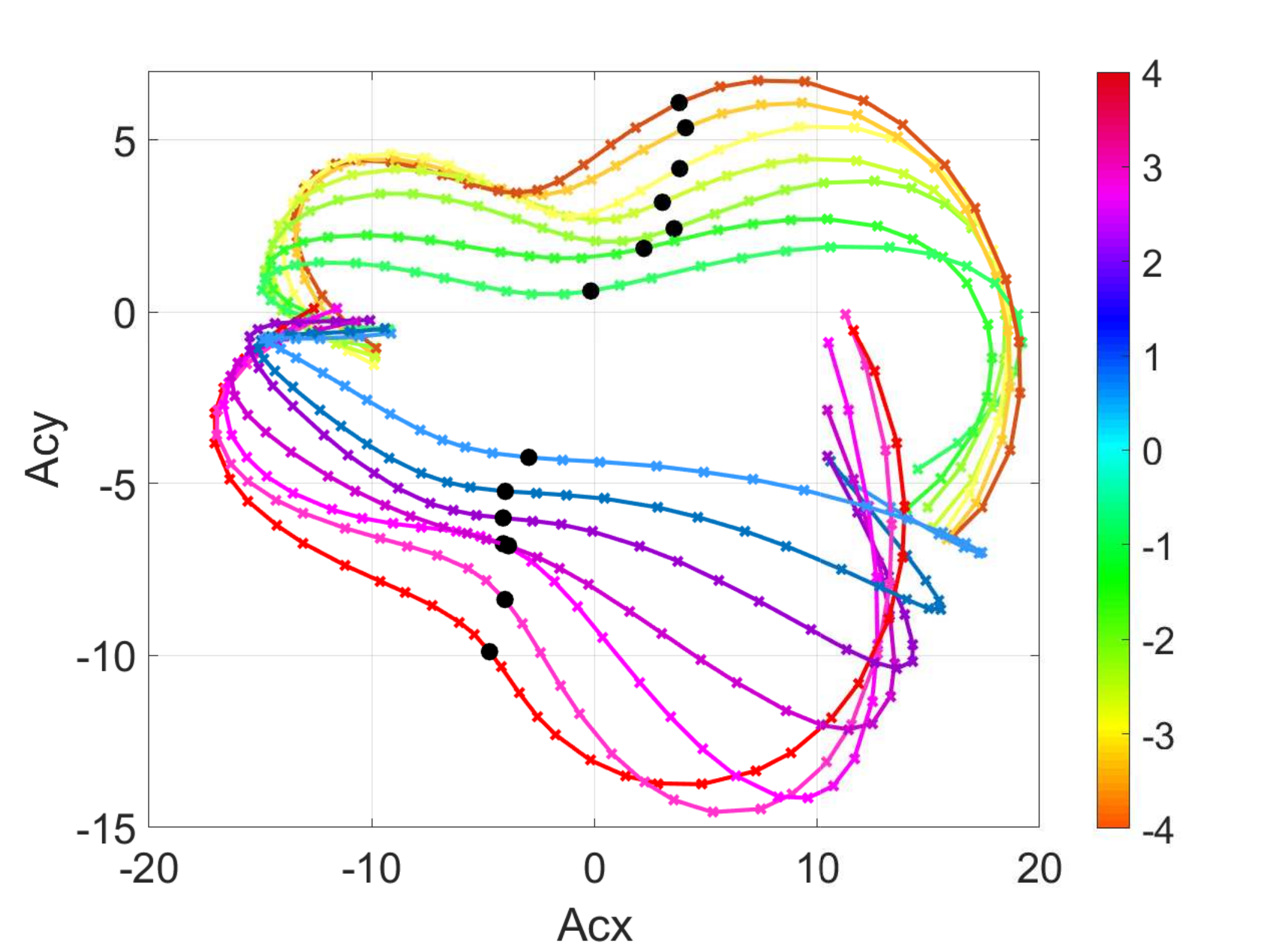} &
\includegraphics[width=0.4\linewidth]{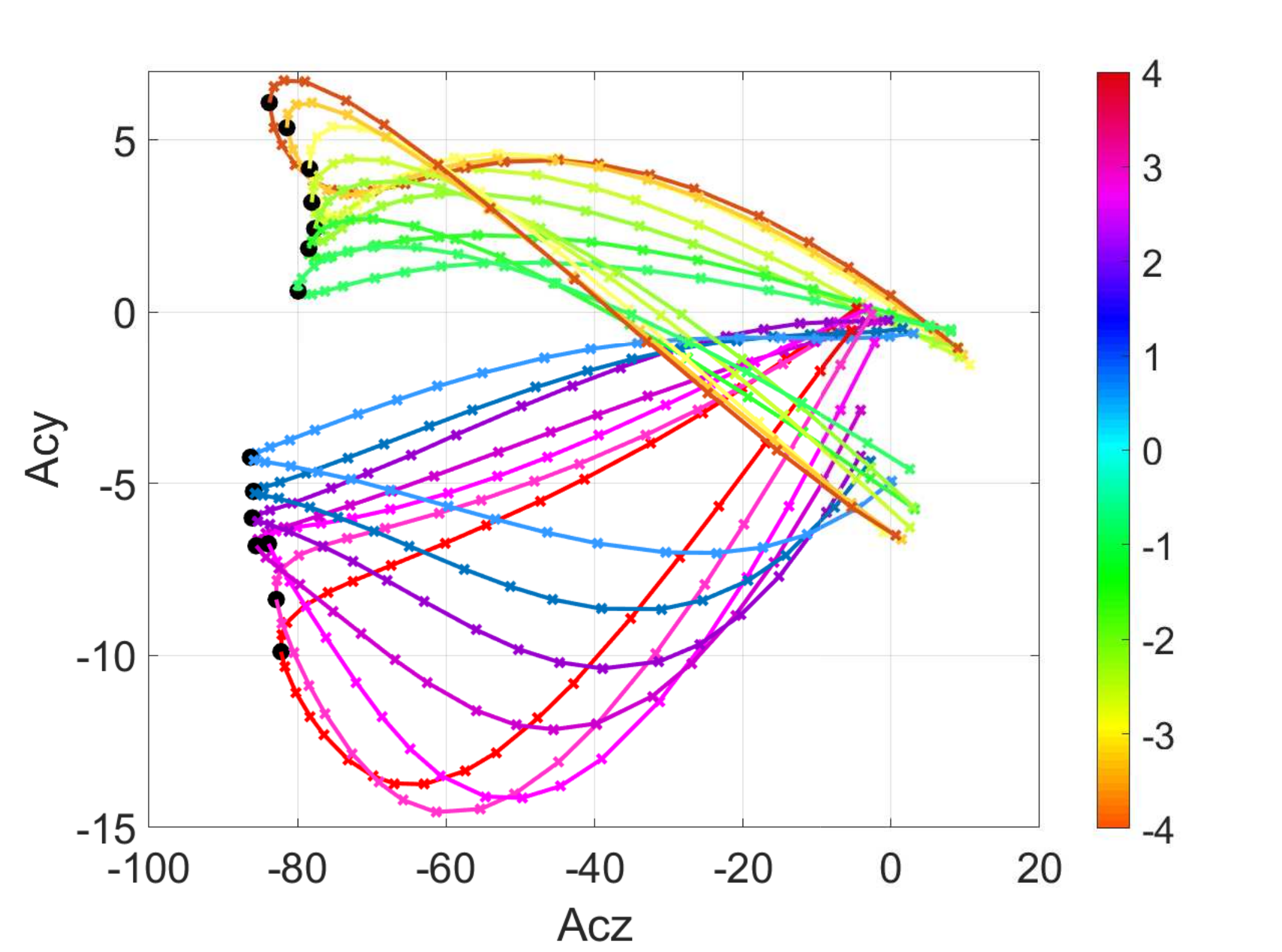}
\end{tabular}
\caption{Mean values of accelerations measurements over low load and high tire speed experiments, for different slip angles (Data set 2), colors represent different slip angles and black dots the minimum $Ac_z$ values for each slip angle.}
\label{fig:training_piezo}
\end{figure*}

Figure \ref{fig:training_piezo} represents the mean values of accelerations measurements over low load and high tire speed experiments for different slip angles with the Data set 2.~Again,~two distinct regions can be found based on negative and positive slip angles. For almost all range of positive slip angles, a negative normal acceleration is collected from the accelerometers.~In addition, the minimum values of normal accelerations in the case of negative slip angles is higher than the cases of positive slip angles.~This might be attributed to the location of accelerometer in the inner liner of tire, as it is not exactly located at the middle of the inner liner.~Another important observation is the stretching of acceleration signals in the positive side of longitudinal accelerations (see $Ac_x-Ac_z$ window).~This can be because of the existence of tire rolling resistance, which results in the generation of unsymmetrical acceleration signals in $x$ direction. The $Ac_y-Ac_x$ window mimics a pear shape, and clearly shows that with increasing the slip angle from -4 degrees to 4 degrees, the value of lateral accelerations decreases (in terms of positive/negative signs).

In addition, two distinct regions can be observed based on the slip angles where the negative slip angles results in positive lateral accelerations, and positive slip angles generates negative lateral accelerations.~This behavior of lateral accelerations generated by accelerometers clearly follows the relation of lateral forces and slip angles.~Indeed, Fig.  \ref{fig:training_piezo} provides an overarching view from the relation between  accelerations in different directions and the corresponding tire forces. Basically, it shows a similar behavior to tire forces that can be observed in acceleration signals considering different slip angles and loads.~Accordingly, the signals presented in Fig.  \ref{fig:training_piezo} can be considered as a framework and basis for the analysis of intelligent tire system with the aim of tire force predictions.

\begin{figure}[!htb]
\centering
\begin{tabular}{c}
(a) \\ \includegraphics[width=0.95\linewidth]{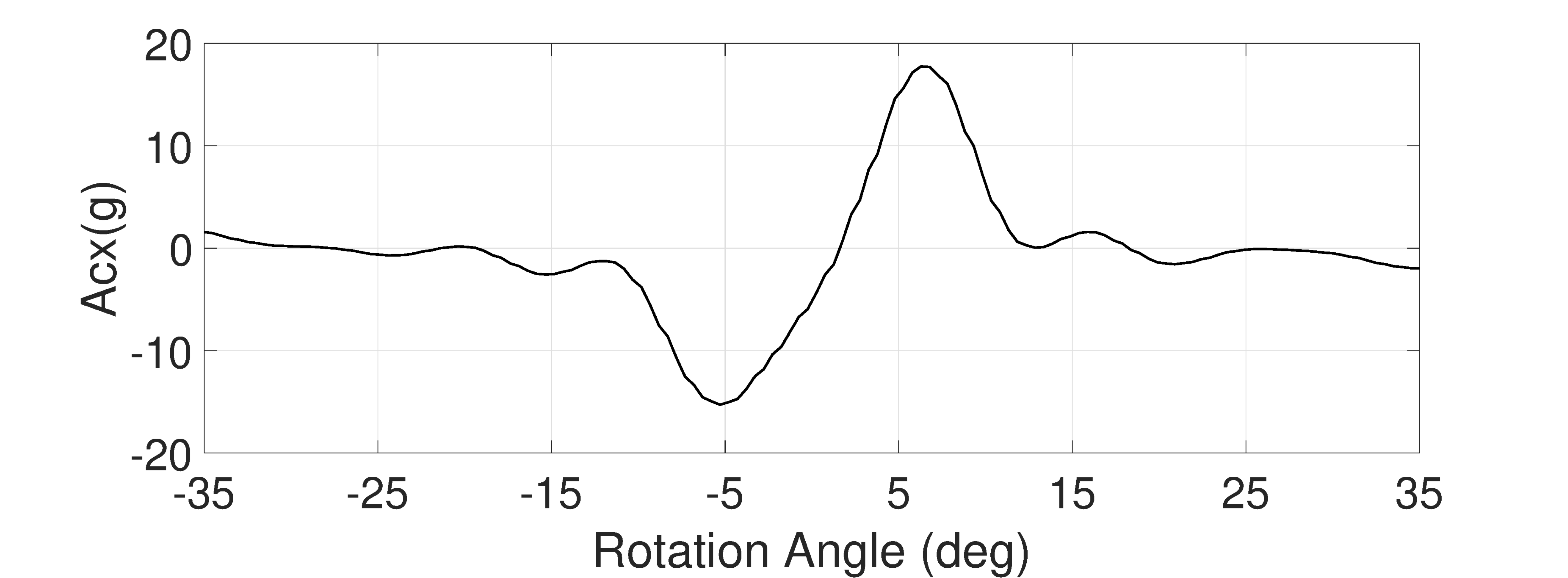} \\ (b) \\
\includegraphics[width=0.95\linewidth]{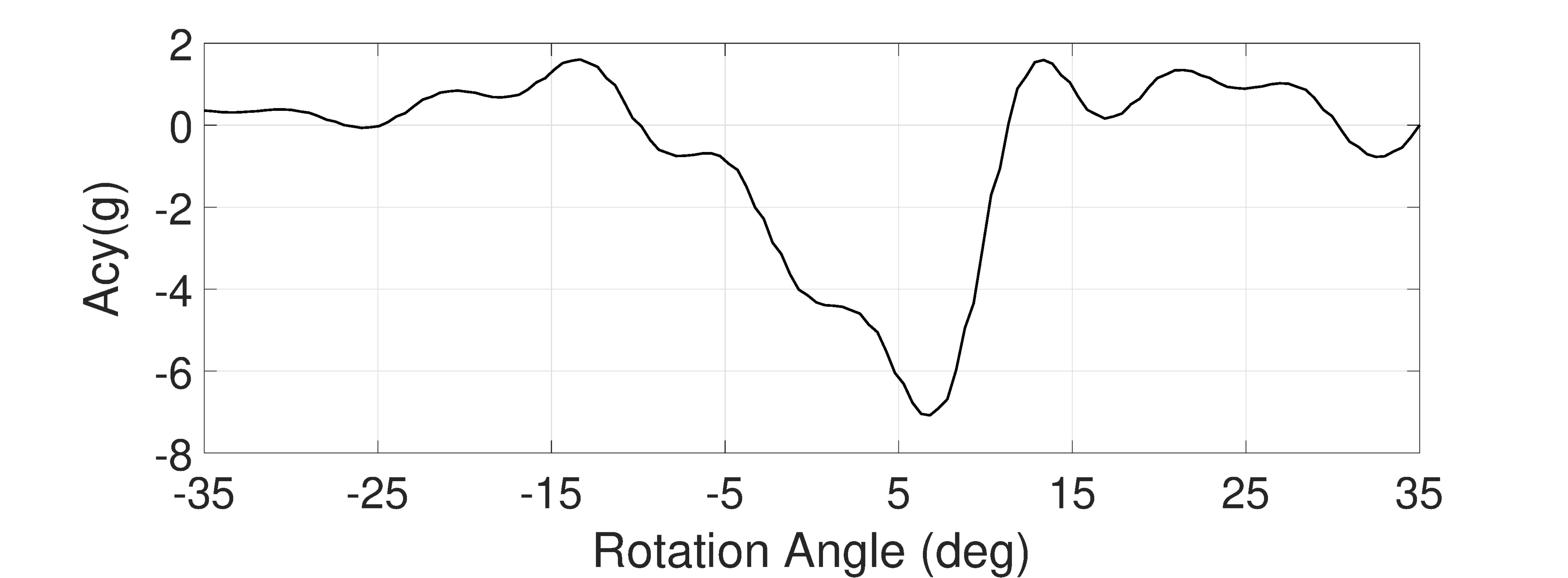} \\
(c) \\
\includegraphics[width=0.95\linewidth]{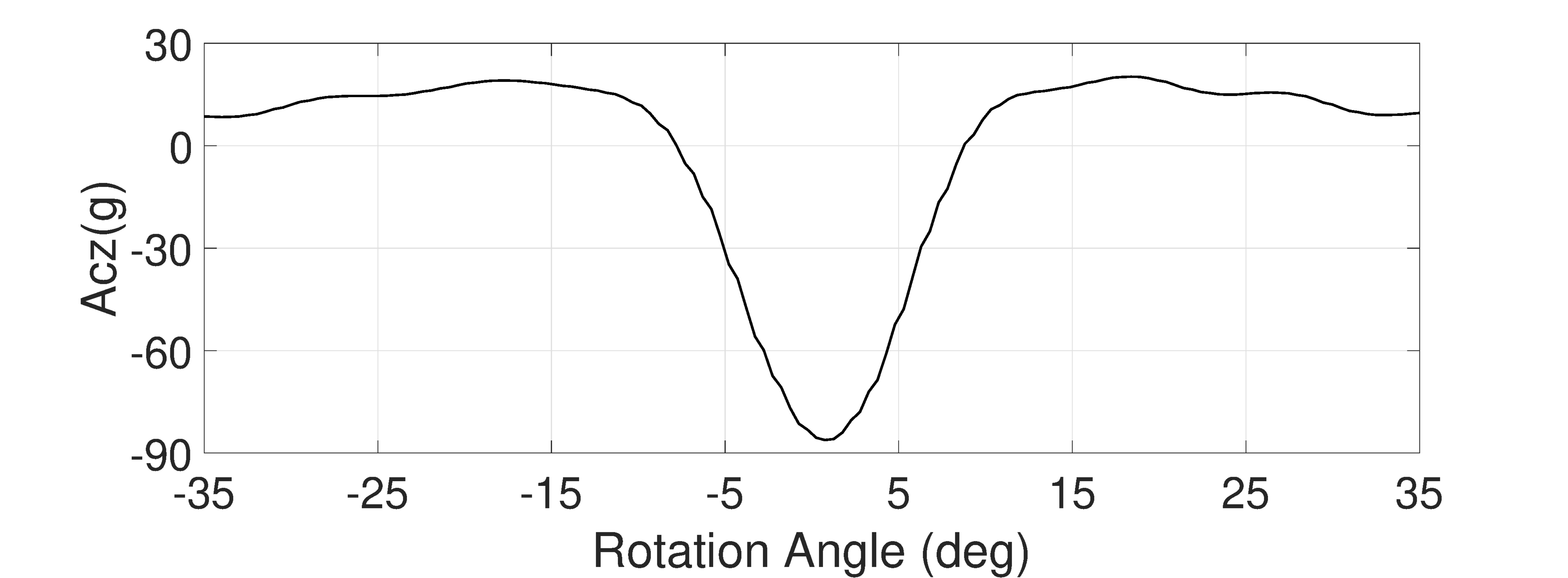} 
\end{tabular}
\caption{Samples of accelerations measurements over one tire rotation for high speed, low load and $1^\circ$ of slip angle (Data set 2): (a) $Ac_x$, (b) $Ac_y$ and (c) $Ac_z$.}
\label{fig:acxyz_abcde}
\end{figure}

Figure \ref{fig:acxyz_abcde}~(a) shows the longitudinal acceleration signals for the  specified angles of -35 to 35 degrees for a full tire rotation. As it enters to the contact patch, a variation of tire radius happens resulting in three different distinct regions.~From the beginning of tire contact patch to the minimum value of longitudinal accelerations presented in  Figure \ref{fig:acxyz_abcde}~(a), we have the first region in which tire radius is larger than the effective radius, and therefore the speed of tire tread is higher than ground velocity.~From the minimum longitudinal acceleration to its peak value, a releasing process occurs in which the tire radius is less than the effective radius so that the velocity in tread is less than the ground velocity. In the third region, which is from the peak longitudinal acceleration to the end of tire contact patch, we have the same trend in radius as the first region described above. Basically, these three regions with three different velocities with respect to the ground result in longitudinal force generation, and accordingly, longitudinal accelerations in accelerometer for the case of pure slip or low slip angles.~It should be noted that the sum of the area under longitudinal accelerations curve is not equal to zero, and this result  manifests the rolling resistance in general.~As the tire experiences low slip angle, an unsymmetrical lateral acceleration is also generated in the accelerometers, see Fig. \ref{fig:acxyz_abcde}~(b). 

Comparing three components of accelerations presented in Fig. \ref{fig:acxyz_abcde} (a-c) delineates that in the case  of low slip angles, the lateral component of acceleration is triggered less than two other components, namely vertical and longitudinal accelerations.~The absolute value of maximum amount of the generated lateral acceleration is almost 7.1\,g comparing to 88\,g and 18\,g for vertical and longitudinal accelerations, respectively.

\section{Lateral Force Prediction}
\label{sec:force}

As illustrated before, the tire lateral force is an important parameter that should be monitored for the vehicle control purposes. Furthermore, besides the lateral force estimation, providing confidence levels for this prediction may be useful for developing vehicle control strategies. Thus, a GPR model is developed in the present work.~In addition, its design procedure and results are presented in this section.

\subsection{Input Variables Selection}

In order to build a mathematical model to estimate the lateral force (desired output) considering the available measured accelerations (input variables), consider that historical data $\mathcal{Z}$ is available such that $\mathcal{Z} \in  \mathbb{R}^{N \times (r+1)}$, where $\mathcal{Z} = [\mathbf{y} \ \ \mathbf{u_1} \ \ \mathbf{u_2} \ldots \mathbf{u_r}]$, $\mathbf{y}$ represents the output force, $\mathbf{u}$ represents the three-axis accelerations measured over the contact patch (interval between points A and E, see Fig.~\ref{fig:contacpatch}), $N$ is the number of samples (tire rotations) and $r$ is the number of measured accelerations used as model's inputs.

\begin{figure}[t]
\centering
\begin{tabular}{c}
(a) \\
\includegraphics[width=0.95\linewidth]{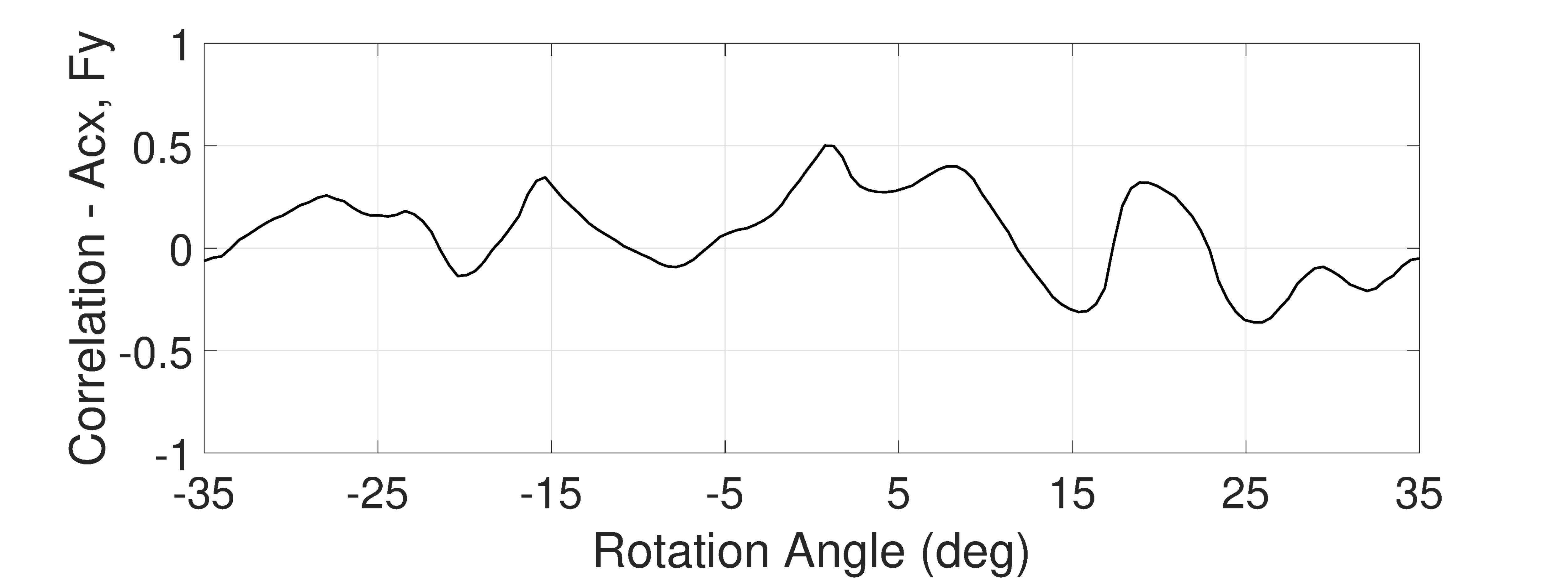} \\ 
 (b) \\
\includegraphics[width=0.95\linewidth]{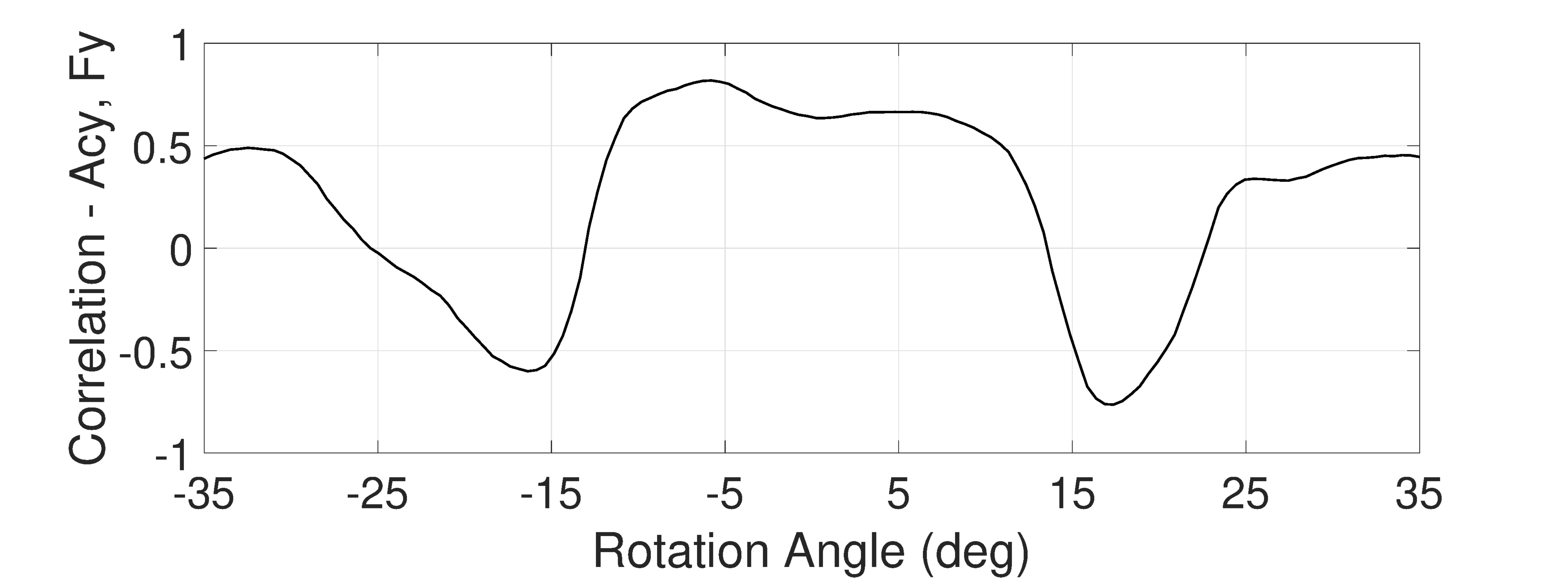} \\
(c) \\
\includegraphics[width=0.95\linewidth]{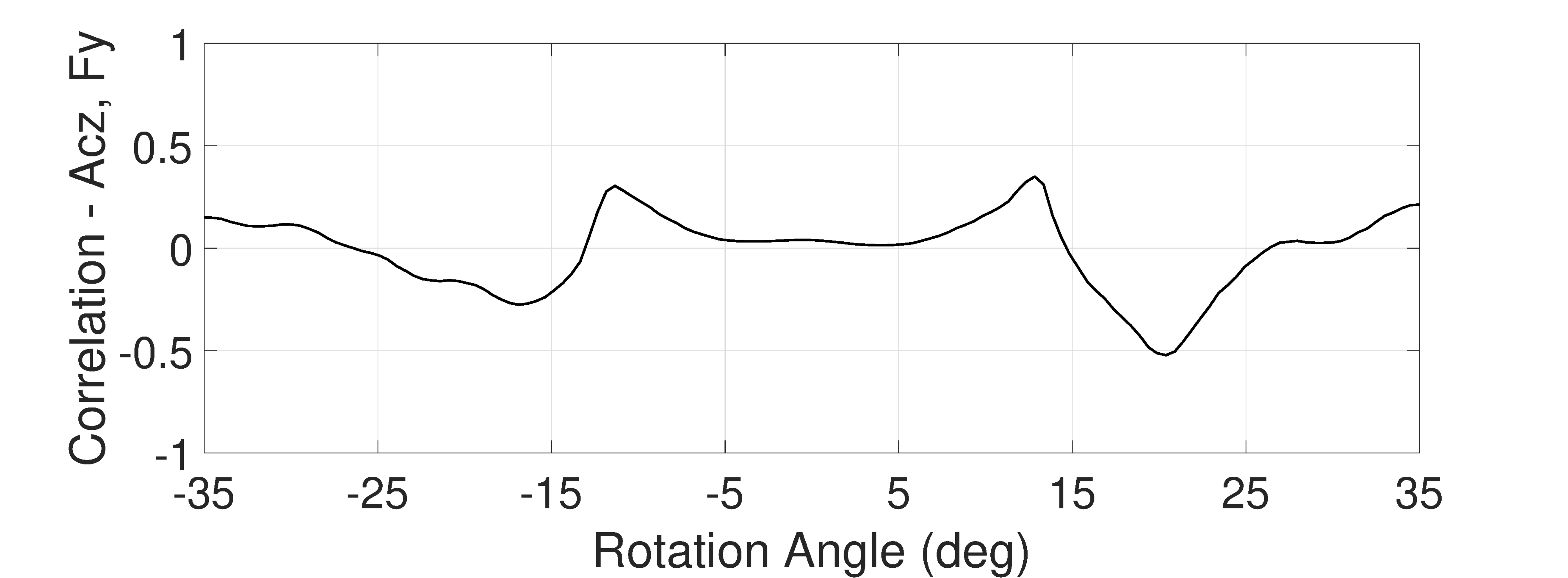} 
\end{tabular}
\caption{Correlation between acceleration measurements and lateral force (Data set 1 with slip angle below $6^\circ$): (a) $Ac_x$, (b) $Ac_y$ and (c) $Ac_z$.}
\label{fig:corr}
\end{figure}

To identify the lateral force prediction model only Data set 1 was employed, that is, Data set 1 is considered the training data set whereas Data set 2 is only used to verify the model generalization (test data). Thus, considering that 912 samples are available in the training data set ($N = 912$) and 420 input variables ($r = 140 \times 3$) are also available (140 measured points for each acceleration direction), an important step to build the model is to define the most relevant inputs. Figure~\ref{fig:corr} presents the Pearson correlation indices between the measured lateral force and each acceleration measurement. As expected, the measured lateral acceleration ($Ac_y$) reached the highest correlation values, although some correlation was also observed for the longitudinal ($Ac_x$) and radial ($Ac_z$) accelerations.

To understand the contribution of the three accelerations components for predicting the lateral force, GPR models were identified using a holdout cross-validation procedure (20 repetitions) over the training data, where 70\% of data were used for training and the remaining 30\% for validation. The GPR selected kernel function was from class \textit{Matérn} with parameter $v=3/2$,  according to \cite{rasmussen2006}, this is normally appropriate for machine learning. Considering the automatic relevance determination (ARD) \cite{rasmussen2006}, Eq. \ref{eq:matern} becomes:
\vspace{0.05cm}

\begin{equation}
   \kappa_{v = 3/2}(\tau) = \sigma_f^2\left( 1 + \sqrt{3}\frac{\tau}{l}\right ) \rm{exp}\left(- \sqrt{3}\frac{\tau}{l}\right ),
\end{equation}
\vspace{0.05cm}

\noindent with positive hyperparameters $\mathbf{\theta} = \{ \sigma_f^2, l\}$, where $\sigma_f^2$ is the output scale and $l$ is the input scale. The hyperparameters is obtained through a gradient-ascent based optimization tool, by maximizing the log-likelihood of the training data:

\begin{equation}
    {\rm log} \ p(\mathbf{y}|\mathbf{X},\mathbf{\theta}) = -\frac{1}{2} \mathbf{y}^{\rm T} \mathbf{K_y}^{-1}\mathbf{y} - \frac{1}{2}|\mathbf{K_y}| - \frac{n}{2}{\rm log} \ 2 \pi, 
\end{equation}
\vspace{0.05cm}

\noindent where $\mathbf{K_y} = \mathbf{K}(\mathbf{X},\mathbf{X}) + \sigma_\epsilon \mathbf{I}$, and the bias-variance trade-off is automatically achieved \cite{liu2020gprscale}.

The prediction error is defined as the normalized root mean square error (NRMSE) as also used in \cite{rezaeian2014novel}: 

\begin{equation}
    {\rm NRMSE} (\%) = 100 \frac{\sqrt{ \frac{1}{N} \sum_{i=1}^N y_i - \hat{y}_i }}{y_{\rm max}},
\end{equation}
\vspace{0.05cm}

\noindent where $\hat{\mathbf{y}}$ is the predicted model output and $y_{\rm max}$ is the maximum value of $\mathbf{y}$. 

\begin{figure}[t]
\centering
\begin{tabular}{c}
(a) \\ \includegraphics[width=0.95\linewidth]{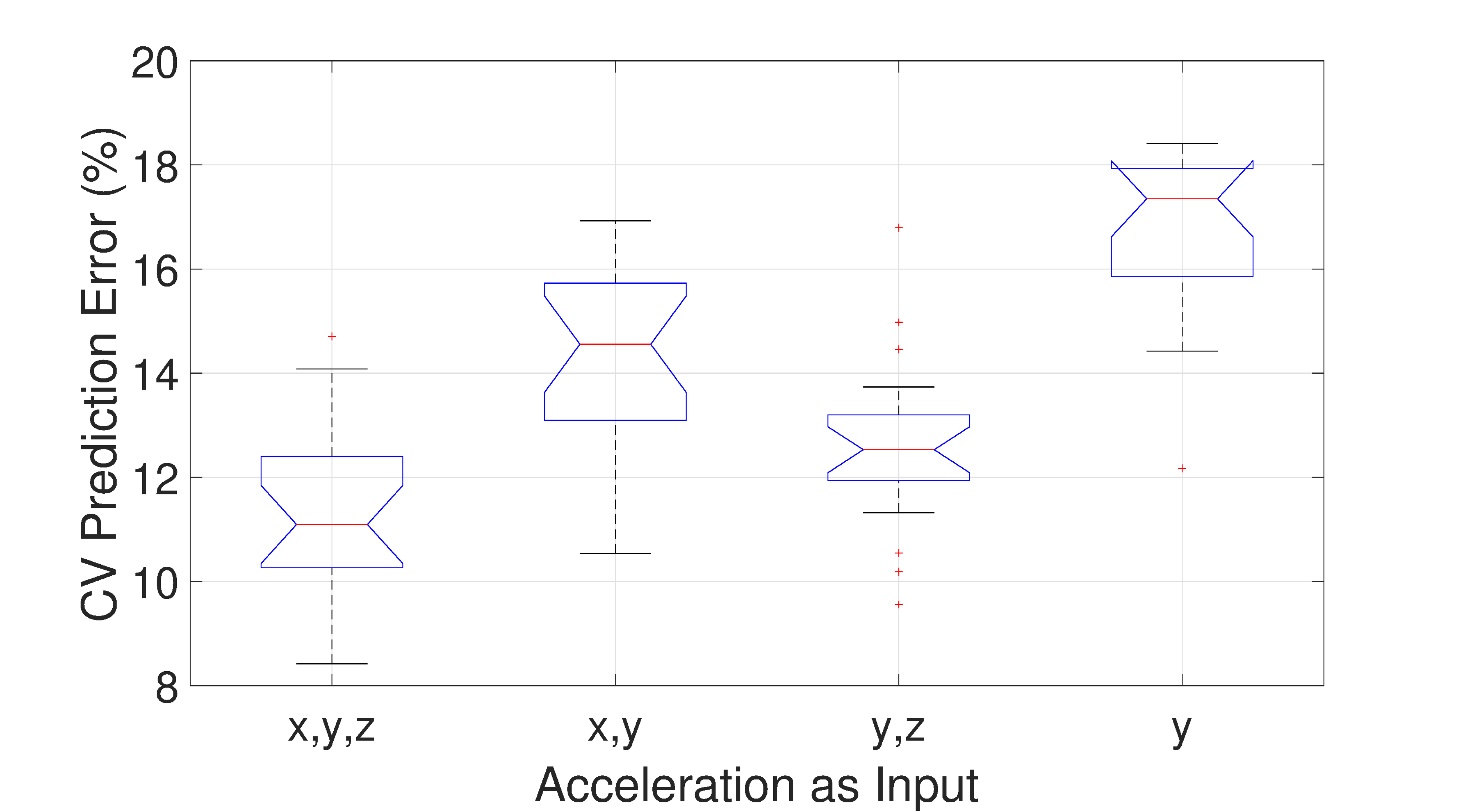} \\ 
 (b) \\
\includegraphics[width=0.95\linewidth]{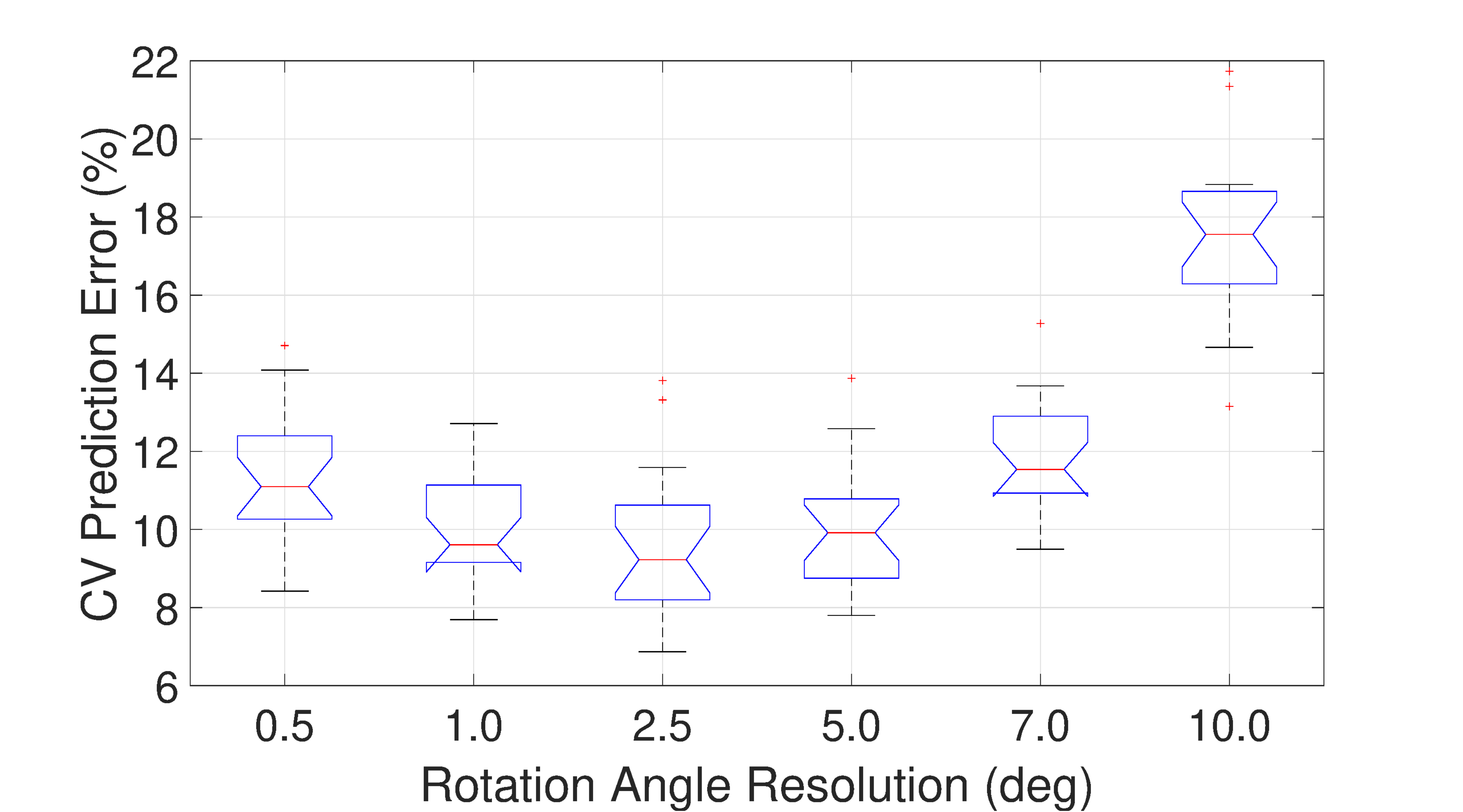} \\
\end{tabular}
\caption{Comparison between GPR performance over Data set 1 (cross-validation): (a) for different acceleration measurements as inputs and (b) different contact patch resolution.}
\label{fig:inputsel}
\end{figure}

The GPR box-plot error results for different configuration of acceleration measurements as inputs are shown in Figure~\ref{fig:inputsel} (a). It is worth noting that the worst result is obtained when only the lateral acceleration measurements are used to predict the lateral force. It means that, besides this is obviously the most important input variable to predict the lateral force, the use of the longitudinal \textit{and} radial accelerations helps to improve the model performance, since the best results were achieved when all three acceleration components over the contact patch were employed.~Another finding is that using both radial and lateral accelerations in the training of GPR provides more accurate results in comparison with the case of training GPR based on longitudinal and lateral accelerations with the purpose of lateral force predictions.~Generally, tires lateral forces are increased with increasing  normal forces. So that, there is a direct relation between these two forces.

Considering that the three acceleration components are useful for predicting the lateral force, the next step was to verify if all 140 selected measurement points for each acceleration direction are indeed important for the model performance. It is conjectured that there are redundant data and finding a parsimonious model is desirable to both improve the generalization capability and to simplify its application in a real system. Thus, with the same aforementioned GPR configuration functions and cross validation procedure, GPR models were identified using different contact patch resolution, where the highest one is obtained using the available 140 points (0.5 degrees). Figure~\ref{fig:inputsel} (b) presents the box-plot errors where it is possible to infer that the use of a lower resolution as 5 degrees does not worsen statistically the models performance. It is important to highlight that, in this case, instead of using 420 input variables, only 42 (14 for each acceleration direction) are employed, drastically reducing the model complexity.

\subsection{GPR Results}

\begin{figure}
 \centering
  \includegraphics[width=1.0\linewidth]{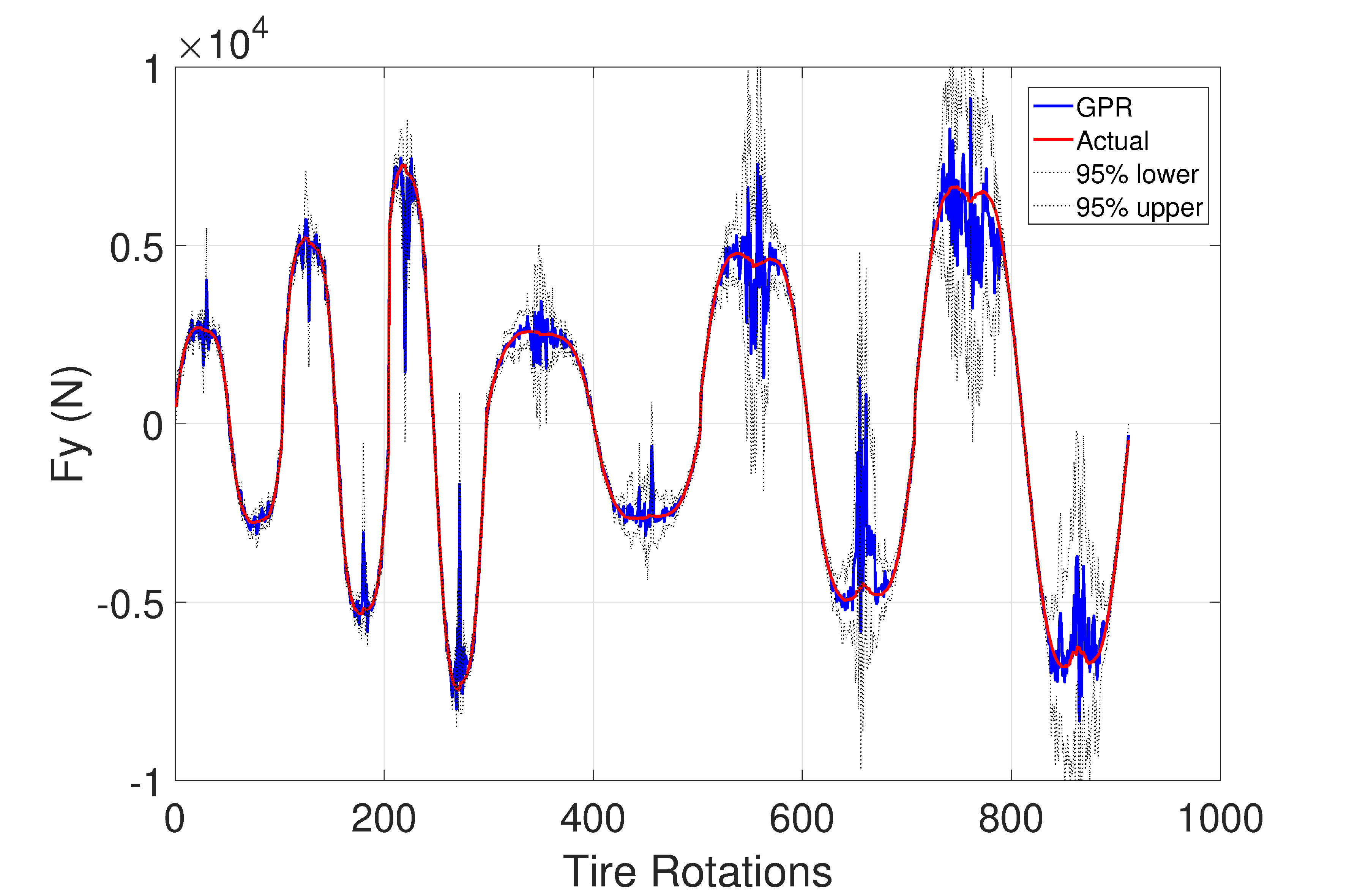}
  \caption{Lateral Force prediction using GPR over validation data ($k$-fold) - Data set 1.}
 \label{fig:gprcv}
\end{figure}

Considering the rotation resolution of 5 degrees achieved promising results, with reduced number of input variables (only 42), this configuration will be further discussed in this section. Firstly, a $k$-fold cross-validation procedure (of 5 folds) was applied to train the GPR model. The use of k-fold procedure here is justified since we aimed to show the validation performance over the training data set, as presented in Fig.~\ref{fig:gprcv}. In this case, since each of the five trained GPR model was validated over disjoint data sets, each lateral force prediction shown in this figure is yielded by only one GPR model. Our observation from Fig.~\ref{fig:gprcv} accentuates that the GPR lateral force predictions are satisfactory over most of Data set 1 and this is always noted for smaller slip angles. However, when high slip angles (typically greater than 6 degrees) and high loads are applied, its performance is worsened. This may be explained by the presence of higher noise level of the acceleration measurements in these situations.

Nonetheless, it is interesting to observe that even in these scenarios, most of the target output is inside the confidence bars provided by the GPR. Besides, these confidence bars can be also used for defining control strategies as discussed in \cite{klenske2016GPRMPCperiodic}.~ The overall cross-validation NRMSE performance of the GPR models on Data set 1 was 9.13\% $\pm$ 1.30\%, which can be considered as a satisfactory value since high values of slip angles are considered in this work.

\begin{figure}[h]
 \centering
\includegraphics[width=1.0\linewidth]{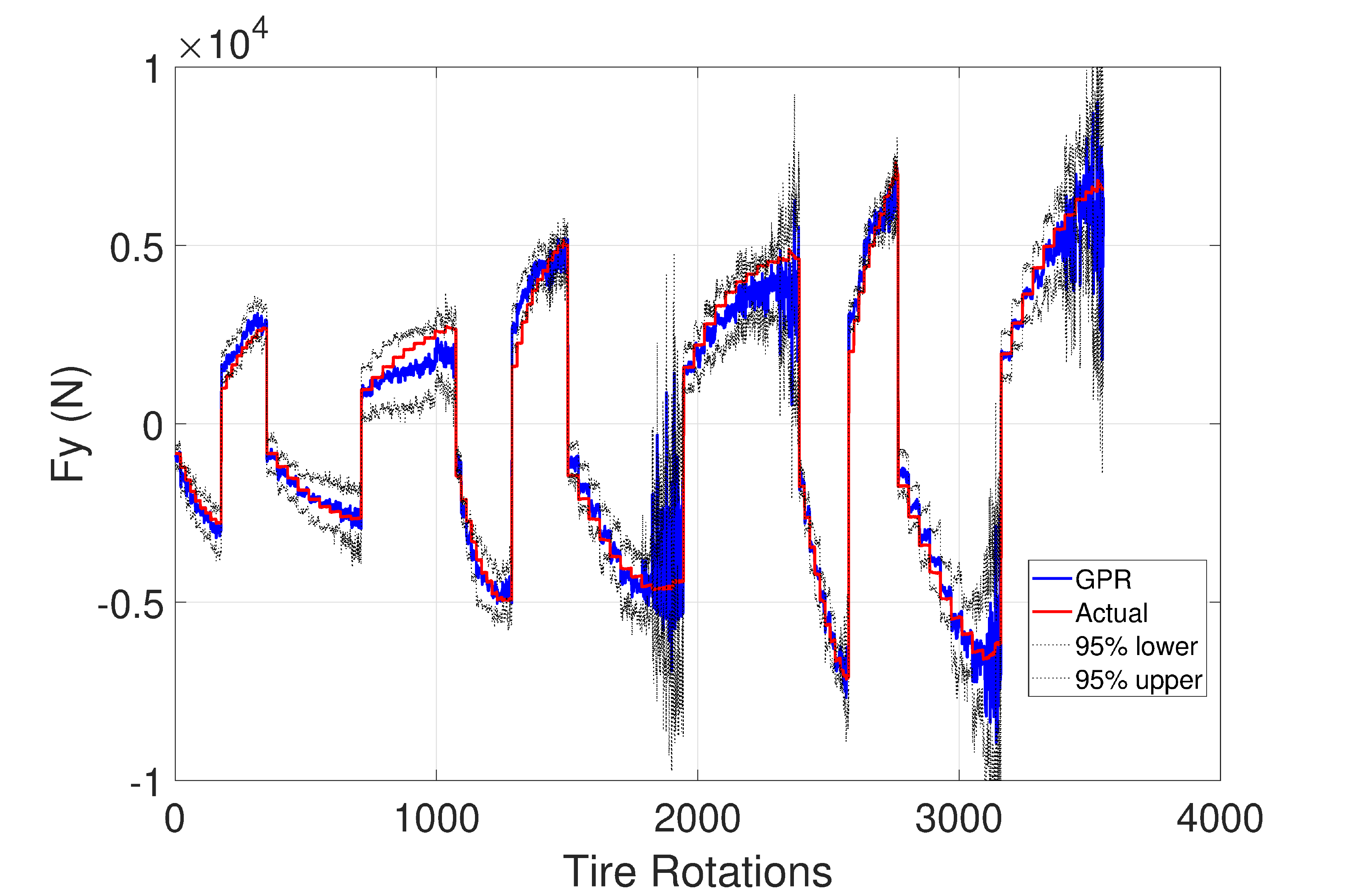}
  \caption{Lateral Force prediction using GPR over the test data set - Data set 2.}
 \label{fig:gprtest}
\end{figure}

\begin{figure}[h]
\centering
\includegraphics[width=0.95\linewidth]{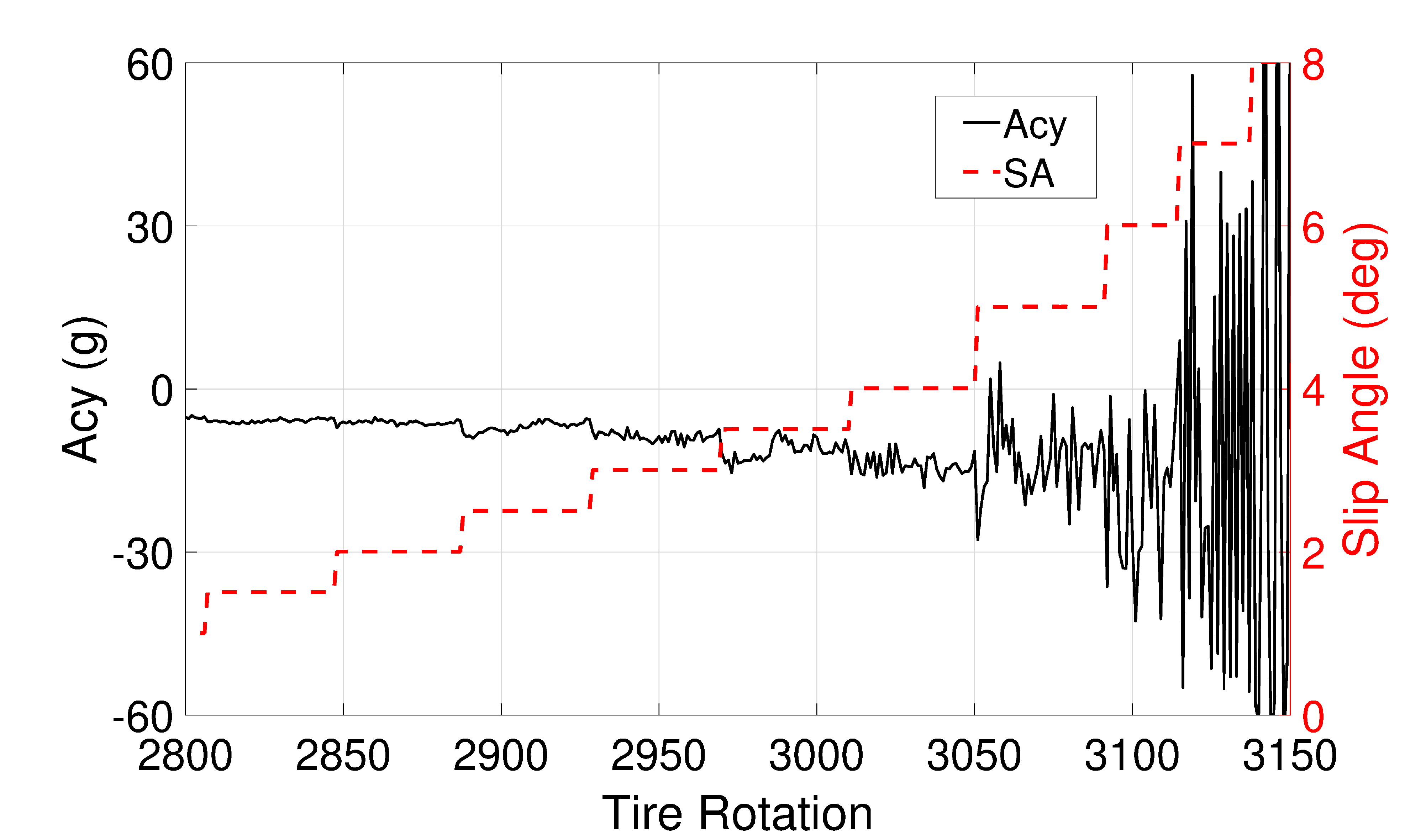} 
\caption{Samples of $Ac_y$ measurements over tire rotations for different slip angles, high speed, high load, and for the contact patch center point (Data set 2).}
\label{fig:acxyz_abcde_sa8}
\end{figure}

\begin{figure}[h]
 \centering
\includegraphics[width=0.95\linewidth]{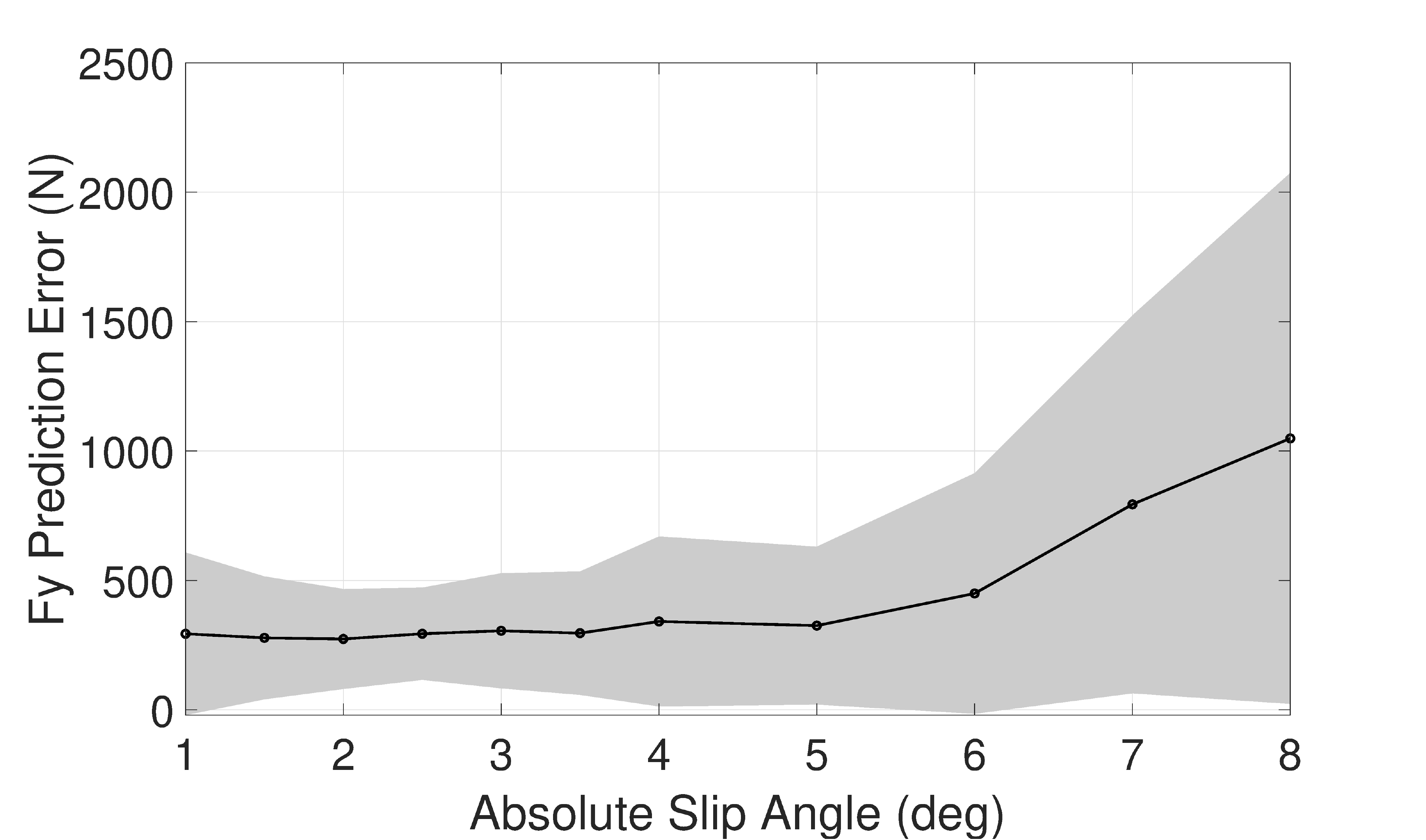} 
  \caption{GPR absolute prediction error analysis (mean and standard deviation) over Data set 2, according to different values of slip angles and with vertical loads from 2080\,N to 6240\,N.}
 \label{fig:error}
\end{figure}

In order to verify the GPR generalization capability, after training the model with all samples of Data set 1, with the same configuration as discussed above, it was tested on Data set 2, which was not used before neither to set nor to select the GPR parameters and functions. The prediction results are shown in Fig.~\ref{fig:gprtest}. The GPR achieved a good prediction accuracy with NRMSE of 7.91\% (R = 0.99), even considering the different kind of maneuvers presented in this data set in relation to the training one. Moreover, it is important to emphasize that, again, the GPR confidence bars comprised the measured lateral forces. As before, the influence of the slip angle and correspondent noisy data on model accuracy can be noticed (see some noisy lateral acceleration measurements samples in Fig.~\ref{fig:acxyz_abcde_sa8}). It is remarkable that GPR lateral force estimation for high slip angles is accompanied with higher prediction uncertainties, showing that the model responded as expected to this challenging situation. To show this influence clearer, the force prediction absolute errors for different values of slip angles are presented in Fig.~\ref{fig:error}, and one can note that the greater the slip angle the higher is the error, mainly when the slip angle is above 6 degrees.

One well-known shortcoming of applying GPR for real-time operation, mainly when dealing with large-scale data sets, is its computational complexity \cite{hewing2019gprMPC}. This is due the calculation of the inverse matrix $[\mathbf{K}(\mathbf{X},\mathbf{X}) + \sigma_\epsilon \mathbf{I} ]^{-1}$ in  Eqs. \ref{eq:gpr1} and \ref{eq:gpr2}, which yields a cost of $\mathcal{O}(n^3)$. Some attempts are presented in the literature to alleviate this burden, which are called scalable GPs, classified into global (e.g. sparse approximations) or local (e.g. mixture of experts) \cite{liu2020gprscale}. Nonetheless, in this paper, the data set used for training the GPR model is not large (912 samples). In order to exemplify the processing time required to predict a sample mean and variance, an execution of the achieved GPR model was carried using MATLAB (version 2019b), on Windows 10 OS with an Intel core $i5-9400$ 2.90 GHz processor and 12 GB of RAM. The mean time required to yield a sample prediction (mean and variance) was equal to  $1.2 \times 10^{-3}$\,s and this should not be a problem for a real-time application. If the prediction time does not satisfy real-time operation requirements due to hardware constraints, the use of scalable GPs is recommended.

\section{Conclusions}
\label{sec:conclusions}

In this paper, an intelligent tire system furnished by an accelerometer was studied.~Applying different kinds of maneuvers, such as varying the normal loads and tire velocity, and also considering high values of slip angle (till 8 degrees), it was shown that the acquired data can underlie interesting analysis about the tire dynamics.~However, it was shown that for high values of slip angles together with the application of high load values, more research is needed to improve  the signal to noise ratio of the accelerometer measurements.~This is probably happening due to severe sliding and high-frequency vibration of the tire in these situations, besides the emergence of a high temperature environment. Further efforts should be made to implement a temperature compensation system to mitigate this problem.

Based on the data extracted through intelligent tire system, a tire lateral force predictor was developed using GPR.~After showing the importance of using the three measured acceleration components for lateral forces estimation, and illustrating that there is an optimal resolution for distributing the measurements points over the tire surface, a good balance between model complexity and accuracy was obtained. The achieved model provides an acceptable generalization performance with illustrating the prediction uncertainties. This is particularly important if some decision making actions were implemented with the predicted lateral forces such as the implementation of MPC controller for vehicle control purposes.~As future work,~the use of GPR can be expanded for the identification of models for other tire variables prediction, for instance, the longitudinal and radial forces besides the tire moments.





\vspace{6pt}

\bibliographystyle{elsarticle-num}
\bibliography{References}

\end{document}